\documentclass{aa}
\usepackage{graphicx}
\usepackage{txfonts}
\usepackage{natbib}
\usepackage{longtable}
\usepackage[shortlabels]{enumitem}
\usepackage{siunitx}

\begin{document}






\title{2D chemical evolution model: the impact of galactic disc asymmetries on azimuthal chemical abundance variations}

\author { E. Spitoni\inst{1}  \thanks {email to: spitoni@phys.au.dk} 
\and  G. Cescutti \inst{2}
\and I. Minchev \inst{3}  \and F. Matteucci\inst{2,4} \and  V. Silva Aguirre
\inst{1} \and \\M. Martig \inst{5} \and G. Bono \inst{6,7} \and C. Chiappini \inst{3}}
\institute{Stellar Astrophysics Centre, Department of Physics and
  Astronomy, Aarhus University, Ny Munkegade 120, DK-8000 Aarhus C
\and I.N.A.F. Osservatorio Astronomico di Trieste, via G.B. Tiepolo
 11, 34131, Trieste, Italy 
\and Leibniz-Institut f\"ur
   Astrophysik Potsdam, An der
  Sternwarte 16, 14482, Potsdam, Germany 
   \and Dipartimento di Fisica, Sezione di Astronomia,
  Universit\`a di Trieste, Via G.~B. Tiepolo 11, I-34143 Trieste,
  Italy
\and Astrophysics Research Institute, Liverpool John Moores
University, 146 Brownlow Hill, Liverpool L3 5RF, UK \and Dipartimento
di Fisica, Universit\`a di Roma Tor Vergata, via della Ricerca
Scientifica 1, I-00133 Rome, Italy
\and I.N.A.F.  Osservatorio Astronomico di Roma, Via Frascati 33, 00078 Monte 
Porzio Catone, Italy
 }

\date{Received xxxx / Accepted xxxx}

\abstract {Galactic disc chemical evolution models generally ignore azimuthal surface density variation that can introduce chemical abundance azimuthal gradients. Recent observations, however, have revealed chemical abundance changes with azimuth in the gas and stellar components of both the Milky Way and external galaxies. }{To quantify the effects of spiral arm density fluctuations on the azimuthal variations of the oxygen and iron abundances in disc galaxies.}{We develop a new 2D galactic disc
chemical evolution model, capable of following not just radial but
also azimuthal inhomogeneities.}{
The density fluctuations resulting from a Milky Way-like  N-body disc formation simulation produce azimuthal variations in the oxygen abundance gradients of the order of 0.1 dex. 
Moreover, in agreement with the most recent observations in
 external galaxies, the azimuthal variations are more evident in
the outer galactic regions.  Using a simple analytical model, we show that the largest fluctuations with azimuth result near the spiral structure corotation resonance, where the relative speed between spiral and gaseous disc is the slowest. }{We provided a new
2D chemical evolution model capable of following azimuthal density
variations.
 Density fluctuations extracted
 from a Milky Way-like dynamical model lead to a scatter in the
 azimuthal variations of the oxygen abundance gradient in agreement
 with observations in external galaxies.   We interpret the presence of azimuthal scatter at all radii by the presence of multiple spiral modes moving at different pattern speeds, as found in both observations and numerical simulations.}

\keywords{Galaxy: evolution, disc -- 
stars: abundances -- ISM: abundances   }

\titlerunning{Impact of density perturbations on the azimuthal gradients}

\authorrunning{Spitoni et al.}

\maketitle

\section{Introduction}

In recent years, integral field spectrographs (IFSs) have largely
substituted long-slit spectrographs in studies designed to characterize
the abundance distribution of chemical elements in external galaxies.
IFSs have permitted for the first time to measure abundances throughout the entire
two-dimensional extent of a galaxy (or a large part thereof) and,
thus, to 
detect azimuthal and radial trends (Vogt et al. 2017).

In the last years,  several   observational  works  have been
found evidence    of significant  azimuthal variations in the abundance gradients in
external galaxies.
S{\'a}nchez et
al. (2015) and S{\'a}nchez-Menguiano et al. (2016) analyzed in detail the chemical inhomogeneities of the
external galaxy NGC 6754 with the Multi Unit
Spectroscopic Explorer (MUSE), concluding that the azimuthal variations of
the oxygen abundances are more evident in the external part of the
considered galaxy.

Vogt et al. (2017) studied the galaxy HCG 91c with  MUSE and
 arrived to the conclusion that the enrichment of the interstellar medium  has proceeded preferentially
along spiral structures, and less efficiently across them.

Azimuthal variations have been detected in the oxygen abundance also
in the external galaxy M101 by Li et al. (2013).  Ho et al. (2017)
presented the spatial distribution of oxygen in the nearby spiral
galaxy NGC 1365.  This galaxy is characterized by a negative abundance
gradient for  oxygen along the disc, but systematic azimuthal
variations of $\sim$ 0.2 dex occur over a wide radial range of
galactic radii and peak at the two spiral arms in NGC 1365.  In the
same work, the authors presented a simple chemical evolution model to
reproduce the observations. Azimuthal variations can be
explained by two physical processes:  after  a  local self enrichment phase in the inter-arm 
region, a consequent mixing 
and dilution phase   si dominant on larger scale (kpc scale) when the spiral density waves pass through.

Probing azimuthal inhomogeneities of
chemical abundances has been attempted in the Milky Way system too.
Balser et al. (2011), measuring H II region
oxygen abundances,  found that the slopes of the gradients differ by a factor of two in their three
Galactic azimuth angle bins.  Moreover, significant local iron abundance
inhomogeneities have also been observed with Galactic
Cepheids (Pedicelli et al. 2009; Genovali et al. 2014).

Balser et al. (2015) underlined the importance of azimuthal
metallicity structure in the Milky Way disc making for the first time
radio recombination line and continuum measurements of 21 HII regions
located between Galactic azimuth $\phi$=90$^{\circ}$- 130$^{\circ}$.
The radial gradient in [O/H] is -0.082 $\pm$ 0.014 dex kpc$^{-1}$ for
$\phi$=90$^{\circ}$- 130$^{\circ}$, about a factor of 2 higher than
the average value between $\phi$=0$^{\circ}$- 60$^{\circ}$.  It was
suggested that this may be due to radial mixing from the Galactic Bar.

Analyzing the Scutum Red-Supergiant (RSG) clusters at the end of the
 Galactic Bar, Davies et al. (2009) concluded that a simple
 one-dimensional parameterisation of the Galaxy abundance patterns
 is insufficient at low Galactocentric distances, as large azimuthal
 variations may be present.  Combining these results with other
 data in the literature points towards  large-scale ( $\sim$ kpc)
 azimuthal variations in abundances at Galactocentric distances of
 3-5 kpc.
It thus appears that the usual approximation of chemical evolution models assuming instantaneous mixing of metallicity in the azimuthal direction is unsubstantiated.

Azimuthal abundance gradients due to radial migration in the vicinity of
spiral arms in a cosmological context have been studied in detail by Grand et
al.  (2012, 2014, 2016),  and  S{\'a}nchez-Menguiano et al. (2016).

Alternatively,
Khoperskov et al. (2018) investigated the formation of azimuthal
metallicity variations in the discs of spiral galaxies in the absence
of initial radial metallicity gradients . Using high-resolution N
-body simulations, they modeled  composite stellar discs, made of
kinematically cold and hot stellar populations, and study their
response to spiral arm perturbations. They found that azimuthal variations in the mean metallicity of stars across a
spiral galaxy are not necessarily a consequence of the reshaping, by radial migration, of an initial radial metallicity gradient. They
indeed arise naturally also in stellar discs which have initially only a negative vertical metallicity gradient.

The aim of this paper is to develop a detailed 2D galactic disc
chemical evolution model, able to follow the evolution of several
chemical elements as in previous 1D models, but also taking into
account azimuthal surface density variations.
In this the paper when we refer to the thin and thick discs we mean
the low- and high-[$\alpha$/Fe] sequences in the [$\alpha$/Fe]-[Fe/H]
plane. Defining the thin and thick discs morphologically, rather than
chemically, identifies a mixture of stars from both the low- and
high-[$\alpha$/Fe] sequences, and vise versa (Minchev et al. 2015,
Martig et al. 2016). It is, therefore, very important to make this
distinction and avoid confusion. 
 We follow the chemical evolution of the thin
disk component, i.e. the low-$\alpha$ population. We assume that the oldest stars of
that low-$\alpha$ component are 
associated with ages of $\sim$ 11 Gyr, in agreement with asteroseismic
age estimates (Silva Aguirre et al. 2018).

 Starting from  the classical 1D Matteucci \&  Fran{\c c}ois
  (1989) approach (the Galactic
disc is assumed to be formed by an infall 
of primordial gas)  we included   2D surface density fluctuation in the
Milky Way disc chemo-dynamical model by Minchev et al. (2013)
(hereafter MCM13), as well as using analytical spiral arm
prescriptions.

Our paper is organized as follows. In Section 2, we describe the framework used for the new model.
In Section 2.1 the adopted nucleosynthesis prescriptions are reported.
In Section 2.2 the density fluctuation from the
chemo-dynamical model by MCM13 are indicated. In Section 2.3 we
present the analytical expressions for the
density perturbations due to
Galactic spiral arm. 
In Section 3 we presents our results with the density fluctuation from
chemo-dynamical models and with an analytical spiral arm prescription are reported.
Finally, our conclusions are drawn in Section 4.

\section{A 2D galactic disc chemical evolution model}
The basis for the 2D chemical evolution model we develop in this section is the classical 1D Matteucci \&  Fran{\c c}ois (1989) approach, 
in which the Galactic
disc is assumed to be formed by an infall 
of primordial gas.
The infall rate  for the thin disc  (the low-$\alpha$ sequence) of a certain
element $i$ at the time $t$ and Galactocentric distance $R$ is
defined as:  
\begin{equation}\label{infall} 
B(R,t,i)= X_{A_i} \, b(R)  \,  e^{-\frac{t}{\tau_D(R)}},
\end{equation}
where $X_{A_i}$ is the abundance by mass of the element
$i$ of the infall gas that here  is assumed
to be primordial,       while 
the quantity $\tau_D(R)$ is the time-scale of gas accretion.
The coefficient $b(R)$ is constrained  by imposing a fit  
to the observed current total surface mass density $\Sigma_{D}$ in the thin disc
as a function of the Galactocentric 
distance given by:
\begin{equation}
\Sigma_D(R,t_G)=\Sigma_{D,0}e^{-R/R_{D}},
\label{mass}
\end{equation}
where $t_G$ is the present time, $\Sigma_{D,0}$  is the central total surface mass density and
$R_{D}$   is the disc scale length. The fit of the $\Sigma_D(R)$ quantity using the infall rate law of eq. (\ref{infall}) is given by: 

\begin{equation}
 \sum_i \int_0^{t_G} X_{A_i} b(R) e^{-\frac{t}{\tau_D(R)}} dt = \Sigma_D (R,t_G),
\end{equation}  

The observed total disc surface mass density  in the solar neighbourhood is
$\Sigma_D (8 \mbox{ kpc}, t_G)=$ 54 M$_{\odot}$ pc$^{-2}$ (see Romano et al. 2000 for a discussion of the choice of this
  surface density).  
The infall rate of gas that follows an exponential law is a fundamental
assumption adopted in most of the detailed numerical chemical
evolution models in which the instantaneous recycling approximation (IRA) is relaxed.

An important ingredient to reproduce the observed radial abundance gradients
along the Galactic disc  is the  inside-out formation on the disc
(Spitoni \& Matteucci 2011, Cescutti et al. 2007, Mott et al. 2013). The timescale $\tau_D(R)$  for the
mass accretion is assumed to increase with the Galactic radius
following a linear relation given by (see Chiappini et al. 2001):

\begin{equation}
 \tau_{D}(R) = 1.033 R(\mbox{kpc}) - 1.27 \mbox{    Gyr}
\label{tau}
\end{equation}
 for Galactocentric distances $\geq$ 4 kpc.
For the star formation rate (SFR) we adopt a Kennicutt (1998) law
 proportional to the gas surface density:
\begin{equation}
  \Psi(R,t) = \nu \Sigma_g^k(R,t),
  \label{SFR}
\end{equation}
 where  $\nu$ is the star formation efficiency (SFE) process and   $\Sigma_g(R,t)$ is the
gas surface density at a given position and time. The exponent
$k$ is fixed to  1.5 (see Kennicutt 1998).

We divide the disc into concentric shells 
1 kpc wide in the radial direction. Each shell is itself divided into 36 segments of width  $\ang{10}$.
Therefore  at a  fixed Galactocentric
distance 36 zones have been created. 

With  this new configuration we can take  into account variations of the SFR along the annular region, produced by density perturbations driven by spiral arms or bars. 
Therefore, an azimuthal
dependence  appears in eq. (\ref{SFR}) and, which can be written as follows:
\begin{equation}
  \Psi(R,t,\phi) = \nu \Sigma_g^k(R,t,\phi).
  \label{SFR2}
\end{equation}

In this paper we will show results related  to the effects of density
fluctuations of the chemo-dynamical model of MCM13 and
we will test the effects of an analytical formulation for the  density
perturbations created by spiral arm waves.
The reference model without any density azimuthal perturbation is similar to the one by Cescutti et al. (2007), which as been shown to be quite successful in reproducing the most recent abundance gradients observed in Cepheids (Genovali et al. 2015).

\subsection{Nucleosynthesis prescriptions}
In this work we present results for the azimuthal variations of
abundance gradients for oxygen and iron.  As done in a number of
chemical evolution models in the past (e.g. Cescutti et al. 2006,
Spitoni et al. 2015, 2019,  Vincenzo et al. 2019), we adopt the nucleosynthesis
prescriptions by Fran{\c c}ois et al. (2004) who provided theoretical
predictions of [element/Fe]-[Fe/H] trends in the solar neighbourhood
for 12 chemical elements.

Fran{\c c}ois et al. (2004) selected the best sets of yields required to best fit the data
 (details related to the observational data collection are in Fran{\c c}ois et al. 2004).
In particular, for the yields of Type II SNe they found that the Woosley
\& Weaver (1995) ones provide the best fit to
the data: no modifications are required for the yields of iron,
as computed for solar chemical composition, whereas for  oxygen,
the best results are given by yields computed as
functions of the metallicity. The theoretical
yields by Iwamoto et al. (1999) are adopted for the Type SNeIa,  and the prescription
for single low-intermediate mass stars is by van den Hoek
\& Groenewegen (1997).

 Although Fran{\c c}ois et al. (2004) prescriptions still
  provide reliable yields for several elements, we must be cautious
  about oxygen.
Recent results have shown that rotation can influence the oxygen nucleosynthesis in massive stars (Meynet \& Meader 2002) and therefore chemical evolution (Cescutti \& Chiappini 2010), in particular at low metallicity. However, this does not affects our results being the data shown in this project relatively metal rich. Moreover, we are mostly interested in differential effects, rather than absolute values.

\subsection{2D disc surface density fluctuations from the MCM13 model}
We consider the  gas density fluctuations present in the Milky Way like
simulation obtained by Martig et al. (2012) and chosen in MCM13 for their
chemodynamical model. 
The simulated galaxy  has a number of properties consistent
with the Milky Way, including a central bar.
MCM13 followed  the disc evolution for a time period
of about 11 Gyr, which is close to the age of the oldest  low-$\alpha$ disc stars
in the Milky Way. 
The classical
1D chemical evolution model is quite successful in reproducing
abundance gradient along the Galactic disc (Cescutti
et al. 2007).

The chemical evolution model used by MCM13 was very similar to the one adopted here; a comparison between its star formation history and that of the simulation was presented in Fig. A.1 by Minchev et al. (2014), showing good agreement.
To extracted the gas density variations we binned the disk into 18
1-kpc-wide radial bins and 10$^{\circ}$-wide azimuthal bins at $|z|<$ 1
kpc. The time resolution is 37.5 Myr for 11 Gyr of evolution. All of
the above is used for our new model described below.

With the aim of preserving  the general trend of the 1D chemical
evolution model, we  introduce a density
contrast function $f$ related to the perturbations originated by the
MCM13 model. At a  fixed
Galactocentric distance $R$, time $t$ and azimuthal coordinate $\phi$,
the new surface mass density is:
\begin{equation}
\Sigma_D(R,t,\phi)=\Sigma_D(R,t) f(\phi,R,t).
\end{equation}
We 
impose that the average value of 
the density contrast $f$ is 1, i.e.:

\begin{equation}
\langle f(\phi,R,t) \rangle_{\phi}=1.
\end{equation}
This guarantees that, at a fixed Galactocentric distance $R$ and a time $t$, the average surface mass density is the one predicted by the 1 D chemical evolution model.

\subsection{ISM density fluctuations from analytical spiral structure}
Here we investigate the effect of an
analytical spiral arm formulation on the azimuthal variations of
the abundance gradients.

  In particular,  we analyse steady wave spiral
   patterns.   As suggested by Bertin et al (1989) and Lin \& Shu
   (1966) when the number of important spiral modes of oscillation is small, the spiral structure is expected to have a highly regular grand design and to evolve in time in a quasi- stationary manner.

 In this work, we consider the model
presented by  Cox \& G{\'o}mez (2002).
The expression for the time 
evolution  of the density perturbation created by spiral arms, referred to an inertial
  reference frame not corotating with the Galactic disk, in terms of the surface mass density is:

\begin{equation}
\Sigma_S(R,\phi,t)= \chi(R,t_G) M(\gamma),
\end{equation}
where $\chi(R,t_G)$ is the present day amplitude of the spiral density:
\begin{equation}
\chi(R,t_G)=\Sigma_{S,0} e^{-\frac{R-R_0}{R_{S}}},
\end{equation}
while
$M(\gamma)$ is the modulation function  for the ``concentrated arms''
given by  Cox \& G{\'o}mez (2002). The $M(\gamma)$ function can be   expressed  as follows:

\begin{equation}
  M(\gamma)= \left(\frac {8}{3 \pi} \cos(\gamma)+\frac {1}{2} \cos(2\gamma) +\frac{8}{15 \pi} \cos(3\gamma)       \right),
  \label{MGAMMA}
\end{equation}

\begin{equation}
  \gamma(R,\phi,t)= m\left[\phi +\Omega_s t -\phi_p(R_0) -\frac{\ln(R/R_0)}{\tan(\alpha)} \right].
  \label{gamma}
\end{equation}
In eq. (\ref{gamma}), $m$  is the number of spiral arms,  $\alpha$ is
the pitch angle, $R_S$ is  the radial scale-length of the drop-off in
density amplitude of the arms,  $\Sigma_{0}$ is the surface arm
density at fiducial radius $R_0$, $\Omega_s$ is the pattern angular
velocity, with the azimuthal coordinate $\phi$ increasing
counter-clockwise and a clockwise rotation,   $\phi_p(R_0)$ is the
coordinate  $\phi$  computed at $t$=0 Gyr and  $R_0$.
An important feature of such a perturbation is that its
average density at a fixed Galactocentric distance $R$ and time $t$  is zero,

\begin{equation}
\langle \Sigma_S \rangle_{\phi}= \Sigma_{S, 0} e^{-\frac{R-R_0}{R_{S}}} \langle M(\gamma) \rangle_{\phi}=0.
\end{equation}

\begin{figure}
 \includegraphics[scale=0.5]{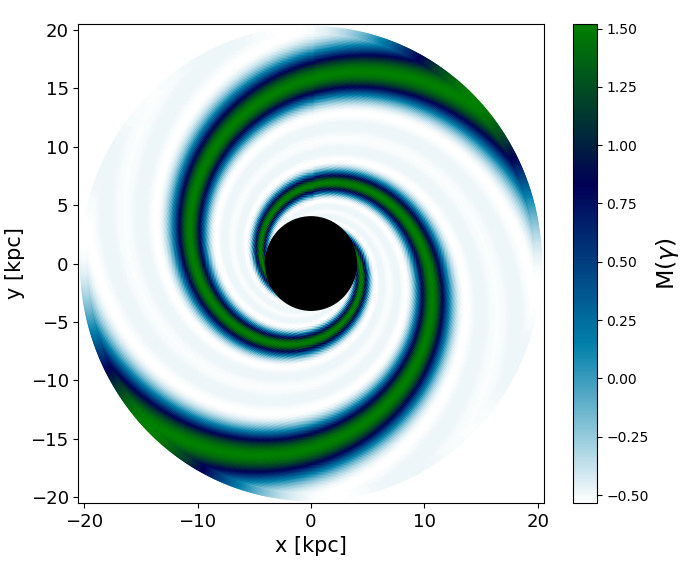}
  \caption{The modulation function $M(\gamma)$ of eq. (\ref{MGAMMA})
    for concentrated arms by Cox \& G{\'o}mez (2002)  with $N=2$ spiral
    arms,  fiducial radius $R_0=8$ kpc,   pitch angle $\alpha=\ang{15}$, and $\phi_p(R_0)=0$.}
\label{MGAMMAF}
\end{figure}

 In Fig. \ref{MGAMMAF} we show the modulation function  $M(\gamma)$ of ``concentrated
arms'' on the Galactic plane  using the  model parameters suggested by Cox \&
G{\'o}mez (2002): $R_0=8$ kpc, $\alpha=\ang{15}$, $R_S=7$
kpc.
 The modulation function is computed at 5 Gyr
assuming the angular velocity value of $\Omega_s$ = 20 km
s$^{-1}$ kpc$^{-1}$ and
$\phi_p(R_0)=0$. 
 In this work we aim  to investigate the effects of spiral arm density
  perturbations on the chemical enrichment  by ejecta from stellar populations
   perfectly corotating with the Galactic
  disk. Our purpose here is the study of regular gas density perturbation
  linked to simple but reliable spiral arm descriptions.

To properly  describe the
  temporal evolution of local density perturbations, the relative  spiral arm speed pattern  
  compared to  the Galactic disk
motion must be computed (further details  will be provided in Section 3.2, in the
Result discussion).

 Cox \& G{\'o}mez (2002) provided a value for the spiral
arm perturbation density at 8 kpc equal to
$\rho_0= \frac{14}{11}$ m$_H$ cm$^{-3}$.
Our implementation requires the surface density $\Sigma_{S,0}$, which
can be recovered from the $z$ direction amplitude provided by
Cox \& G{\'o}mez (2002) 
(their eq. 1), with the following relation:

\begin{equation}
\Sigma_{S,0}=2 \rho_0 \int_0^\infty  \mbox{sech}^2\left(\frac{x}{H}\right)dx=2 H \rho_0,
\end{equation}
where $H$ is  the disc scale-height.
Adopting $H$=180 pc (chosen to match the scale-height of the thin
stellar disc proposed by Dehnen \& Binney 1998, Model2; and in agreement with Spitoni et al. 2008) we obtain

\begin{equation}
\Sigma_{S,0}=21.16 \mbox{  M}_{\odot} \mbox{  pc}^{-2}.
\end{equation}

It is important to underline that in our approach the time dependence
of the density perturbation by the spiral arms is only in the
modulation function $M(\gamma)$ through the term $\Omega_s t$ (see
eqs. \ref{MGAMMA} and \ref{gamma}). Currently, there are no
analytical prescriptions for the time evolution  of both the amplitude 
of the spiral arm perturbation and  its radial profile 
in the  Galactic evolution context (spiral arm  redshift evolution).
 Therefore,   we  make the  reasonable assumption that during the
Galactic evolution  the ratio between the amplitude of the spiral
density perturbation $\chi(R,t)$  and the total surface density
$\Sigma_D(R,t)$ computed at the same  Galactic distance $R$
remains  constant in time, i.e.  $ \frac{d}{dt} \, \left[
\chi(R,t)/\Sigma_D(R,t) \right]$=0, assuming a coeval evolution of
these two structures in time. 
We define the dimensionless  quantity  $\delta_S(R,\phi,t)$ as the  following ratio:

  \begin{equation}
    \delta_S(R,\phi,t)= \frac{ \Sigma_S(R,\phi,t)+ \Sigma_D(R,t)}{\Sigma_D(R,t)}=1 + \frac{ \Sigma_S(R,\phi,t)}{\Sigma_D(R,t)}.
    \label{delta}
  \end{equation}
With the assumption  that the ratio $\chi(R,t)/\Sigma_D(R,t)$ is constant in time, eq. (\ref{delta}) becomes:

  \begin{equation}
    \delta_S(R,\phi,t) =1 + M(\gamma)\frac{
      \chi(R,t_G)}{\Sigma_D(R,t_G)}.
  \label{delta2}
   \end{equation}
If we include the contribution of the perturbation originated by spiral arm in the SFR driven by a linear Schmidt (1959) law (i.e. $\Psi=  \nu \Sigma_g(R,t)$) we have that:  
  
\begin{equation}
  \Psi(R,t,\phi)_{d+s} = \nu \Sigma_g(R,t) \delta_S(R,\phi,t).
  \label{SFR_D}
\end{equation}
We are aware that this is a simplification to the more complex
behavior seen in N-body simulations (Quillen et al. 2011, Minchev et
al. 2012b, Sellwood and Carlberg 2014) and external galaxies
(Elmegreen et al. 1992; Rix \& Zaritsky 1995; Meidt et al. 2009),
where multiple spiral patterns have been found.
 We will make use of this description in Section 3.2.2, where we will consider the simultaneous perturbation by a number of spiral patterns moving at different angular velocities.

\begin{figure*}
\centering
  \includegraphics[scale=.48]{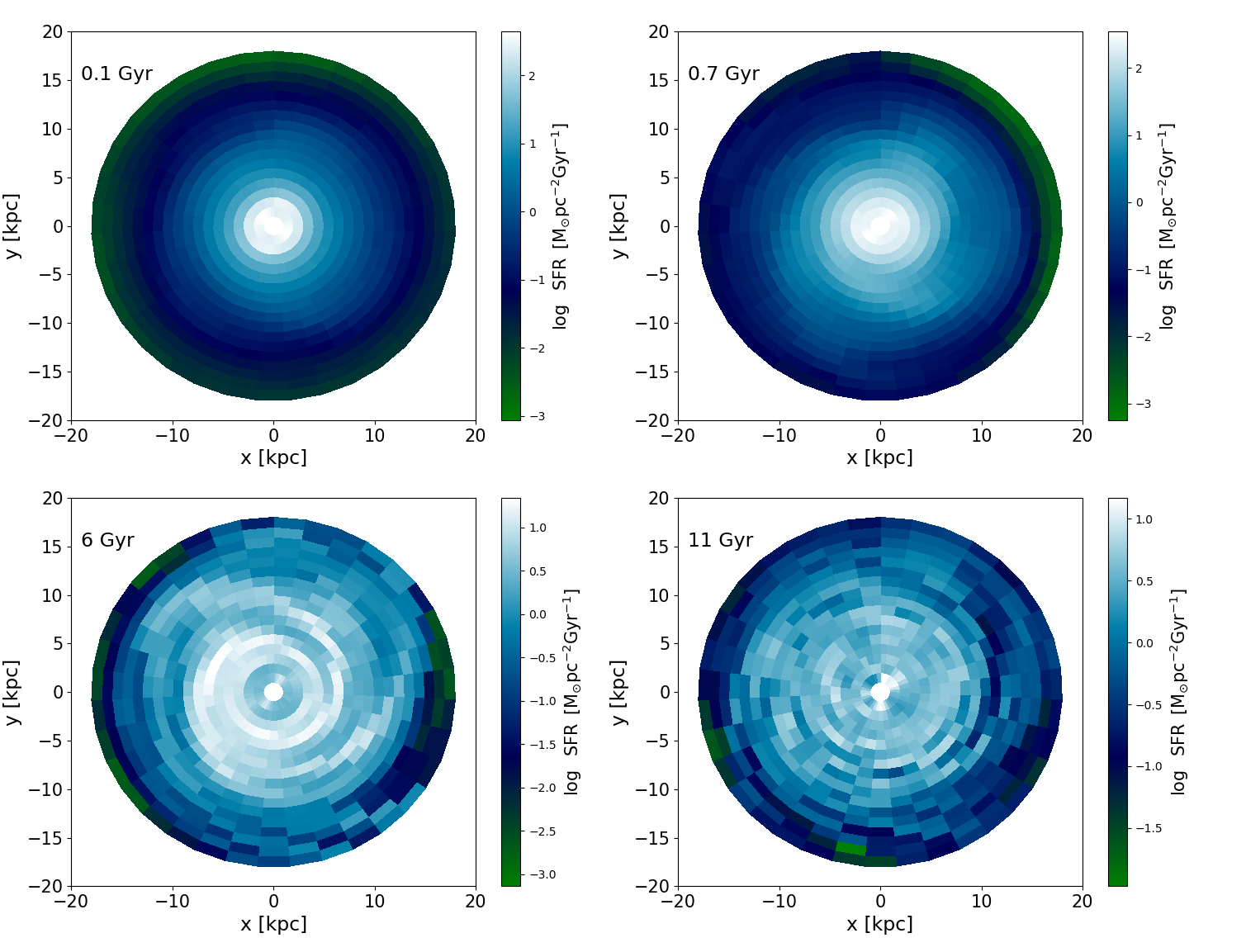}
  \caption{The galactic disc SFR in units of M$_{\odot}$ pc$^{-2}$
    Gyr$^{-1}$ computed at 0.1, 0.7, 6, 11 Gyr after the start of disc formation,  for the
    chemical evolution  model in which we tested the effects of 
the density fluctuations  resulting from the MCM13 model.}
\label{mSFR}
\end{figure*}
\begin{figure*}
\centering
  \includegraphics[scale=.48]{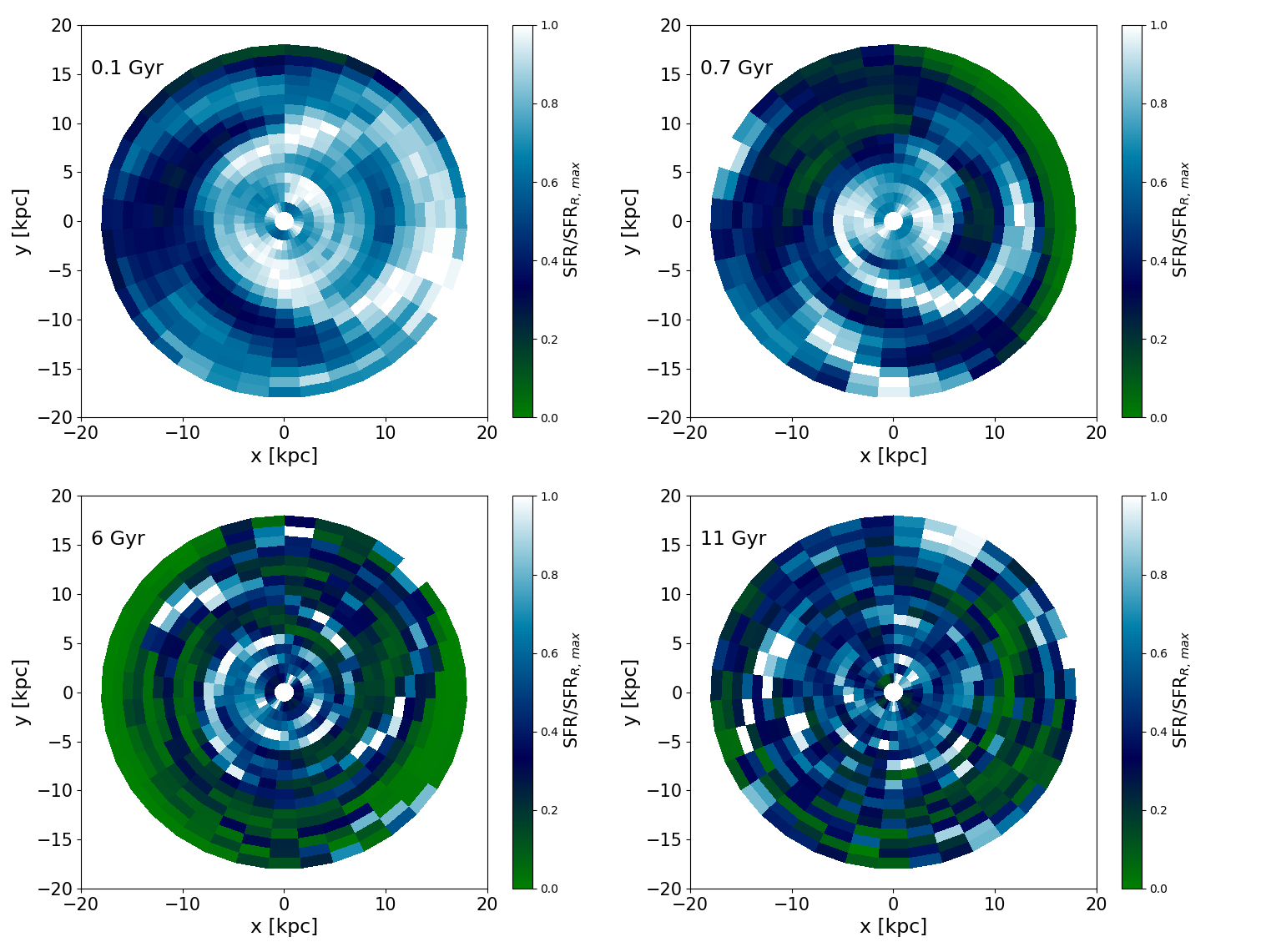}
  \caption{
The galactic disc SFR normalized to the maximum value
    SFR$_{R, ,\ max}$ of the annular region located at the Galactocentric distance
    $R$, i.e SFR($R$, $\phi$)/SFR$_{R,   max}$,
    computed  computed at 0.1, 0.7, 6, 11 Gyr after the start of disc formation,  for the
    chemical evolution  model in which we tested the effects of 
the density fluctuations by MCM13 model.}
  \label{1SFR}
\end{figure*}

As stated in the previous Section, the average modulation function
over the azimuth $\phi$ at a fixed time $t$ and Galactocentric
distance $R$ is null ($\langle M(\gamma) \rangle_{\phi}=0$). Therefore, in presence
of a linear Schmidt (1959) law  at a fixed Galactocentric
distance the average value of $\Psi(R,t,\phi)_{d+s}$ over $\phi$ of
the SFR defined in eq. (\ref{SFR_D}) is equal to the unperturbed SFR
by the following expression:

\begin{displaymath}
 \langle\Psi(R,t,\phi)_{d+s}\rangle_{\phi} =\Psi(R,t) \langle 1 + M(\gamma)\frac{ \chi(R)}{\Sigma_D(R,t_G)}\rangle_{\phi}=
\end{displaymath}
\begin{equation}
 =\Psi(R,t)\left(1+\langle M(\gamma) \rangle_{\phi}\frac{ \chi(R)}{\Sigma_D(R,t_G)}\right)=\Psi(R,t).
\end{equation}
 Here, we do not adopt a linear  Schmidt (1959) law, and we
  use  the
 SFR  proposed by Kennicutt (1998) which exhibits the
exponent $k$=1.5.  Hence, the SFR in the Galactic disc in presence of  spiral arm
density perturbations becomes:
\begin{equation}
  \Psi_k(R,t,\phi)_{d+s} = \nu \Sigma_g(R,t)^k \delta_S(R,\phi,t)^{k}.
  \label{SFR_k}
\end{equation}

 Roberts (1969) provided the  exact shape of the steady gas
distribution in spiral arms, finding an offset between the maximum
of the stellar spiral arm and the maximum of the
gas distribution driven by galactic shocks. In his Figure 7, it is shown
that  
 the regions of newly born luminous stars and the HII regions lie on
the inner side of the observable gaseous spiral arm of HI. 
The presence of  a small but noticeable offset between the gas and
stellar spiral arms  has been
also found   in  the study of interactions between disc galaxies and
perturbing companions in 3D N-body/smoothed hydrodynamical numerical
simulations   by Pettitt  (2006).

Because of uncertainties related to the real magnitude of this
offset (small offsets  are predicted by Pettitt 2006),  in our work we do not consider it, and the SFR is
more enhanced in correspondence   of  the  total density perturbation peak
(see eq. \ref{delta2} and the modulation function in Figure 1).  We are aware
that is true only near  the corotation radius,
however with
our simpler approach we provide an upper limit estimate for the azimuthal
abundance variations generated by steady spiral arms density
perturbations.
In presence of an  off-set the density perturbation should be less ``concentrated'' and
more smeared. 

Our model in the presence of analytical spiral arms  must be considered as
a first attempt to include spiral
 structure in a classical chemical evolution model.  
As stated in Section 2.2, we will also present results for the
 azimuthal abundance variations originated by chemodynamical Milky Way
 like simulation in the presence of spiral arms and bar in a self
 consistent way.  Our analytical spiral arms model is meant to break
 down the problem to understand the reason for the causes of azimuthal
 variations.  Assuming that modes add linearly, we can approximate a
 realistic galactic disk by adding several spiral sets with different
 pattern speeds, as seen in observations (e.g., Meidt et al. 2009) and
 simulations (e.g., Masset \& Tagger 1997, Quillen et al. 2011,
 Minchev et al. 2012a).  
\begin{figure*}
\centering
 \includegraphics[scale=0.58]{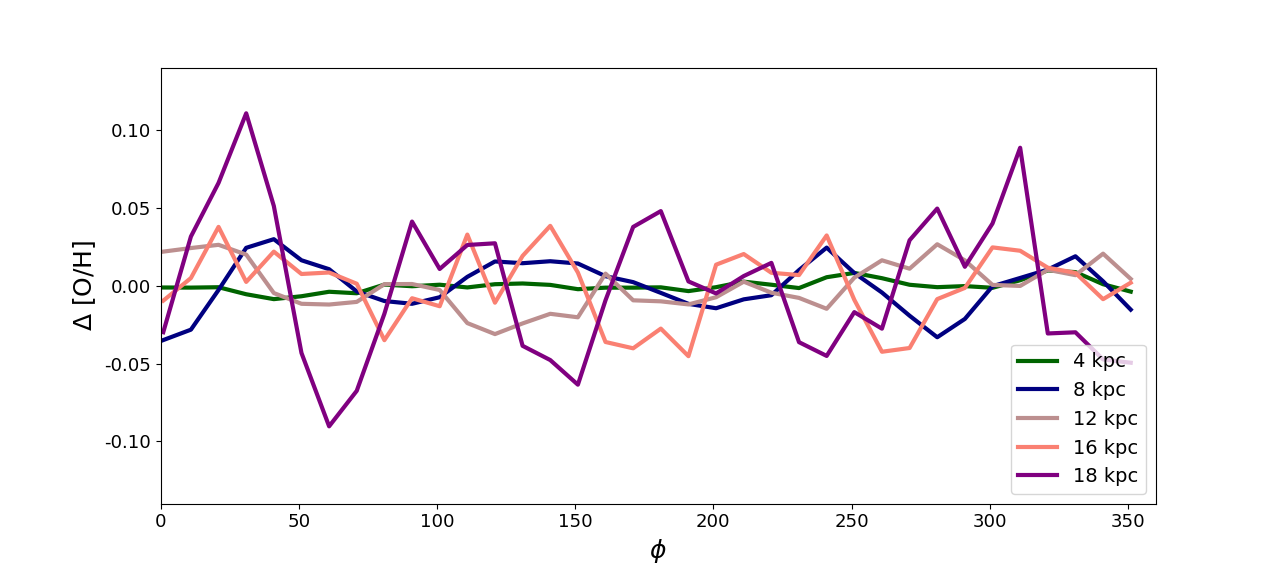}
 \includegraphics[scale=0.58]{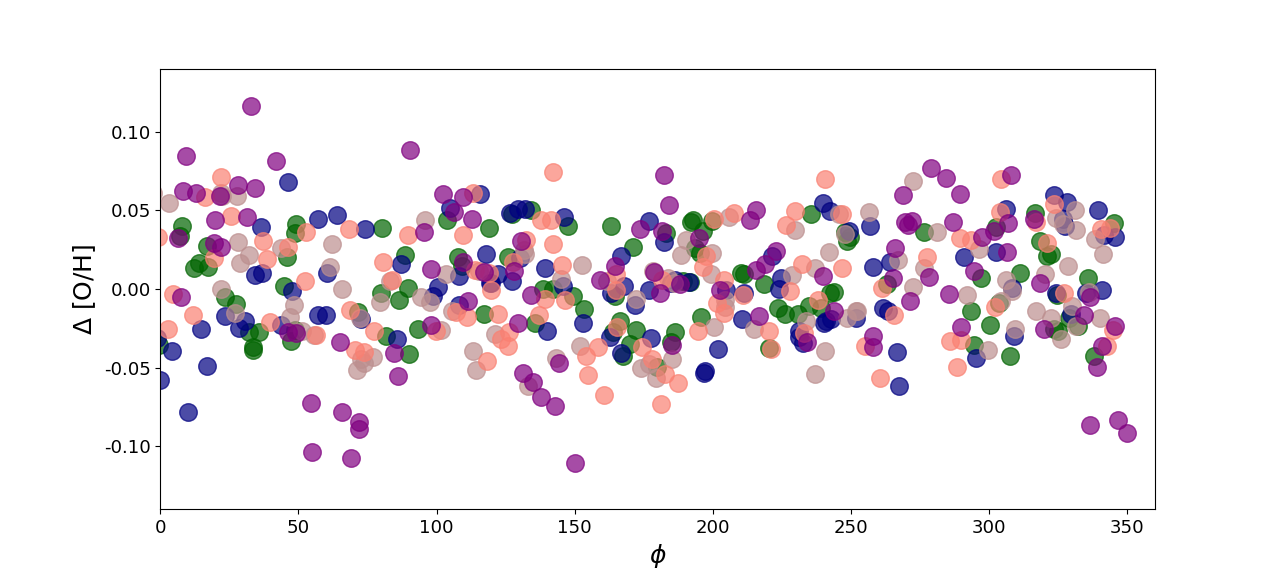}
  \caption{Results for 
the chemical evolution model in which we consider the density
fluctuation by the chemo-dynamical model by MCM13.
  {\it Upper Panel}:  residuals of the ISM oxygen abundances as a function of Galactic azimuth computed with our chemical evolution model at 4, 8, 12, 16, and 18 kpc after subtracting the average radial gradient. 
 {\it Lower Panel}: our mock observation
to mimic S{\'a}nchez et al. (2015) results, in which we randomly plot
residual ISM oxygen
abundance  predicted by our chemical evolution at 4, 8, 12, 16, and 18 kpc
adding an uncertainty of 5$^{\circ}$ in the azimuthal component and
taking into account [O/H] errors according to  S{\'a}nchez et al. (2015); the
color code is identical  to line colors of the upper panel: innermost
disc regions are associated with the green points, the outermost ones with the purple points.   }
\label{minchev_av}
\end{figure*}

\begin{figure*}
\centering
\includegraphics[scale=0.45]{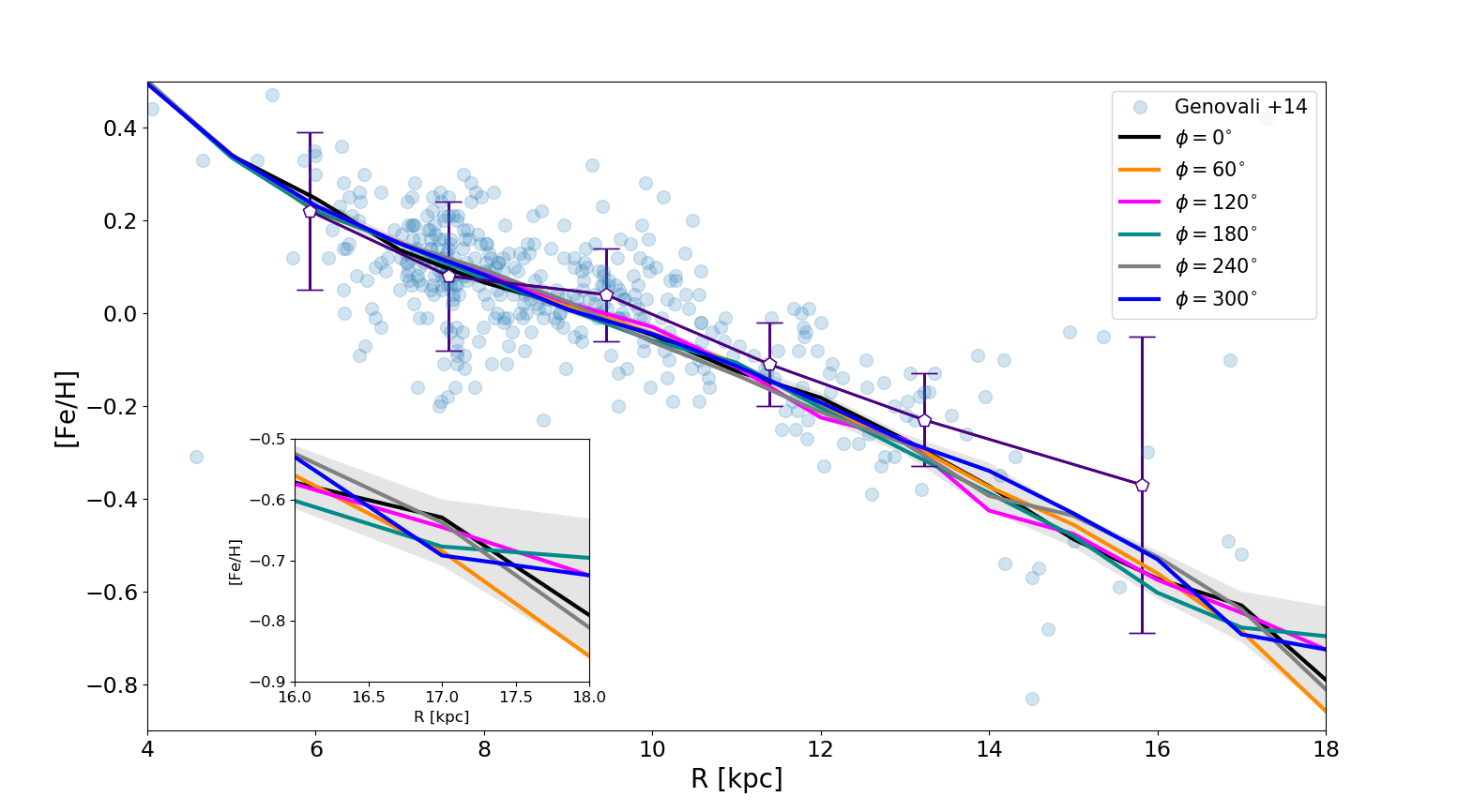}
   \caption{Results for 
the chemical evolution model in which we consider the density
fluctuation by the chemo-dynamical model by MCM13.
The present day  Fe abundance gradient computed at
 different azimuthal coordinates. The shaded grey area limits are
 related to the maximum and minimun iron abundance values at the
 different Galactocentric distances. Observational data (light blue circles)
 are the Cepheids collected  by Genovali at al. (2014).
With the empty pentagons we report the average abundance values 
and relative errors of Genovali et al. (2014) when  divided into six radial
bins. In  the  zoomed region are presented   the
model lines computed between 16 and 18 kpc.
 }
\label{minchev_grad}
\end{figure*}

\begin{figure}
 \includegraphics[scale=0.47]{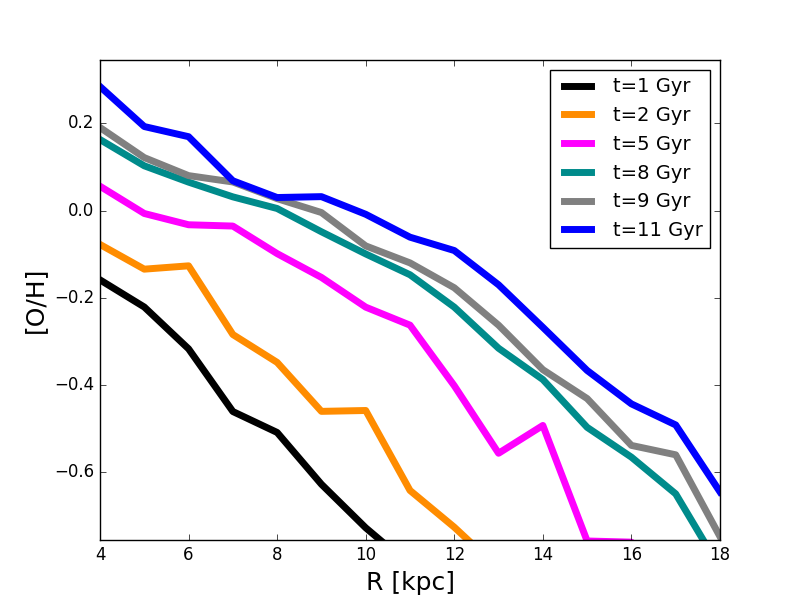}
   \caption{Results for 
the chemical evolution model in which we consider the density
fluctuation by the chemo-dynamical model by MCM13.
Time
evolution of the oxygen abundance gradient at $\phi$=0$^{\circ}$.   }
\label{minchev_grad_ev}
\end{figure}

\begin{figure}
  \includegraphics[scale=0.56]{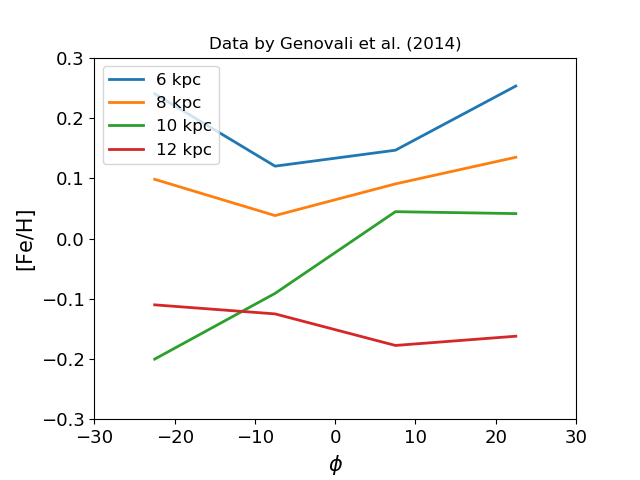}
  \includegraphics[scale=0.40]{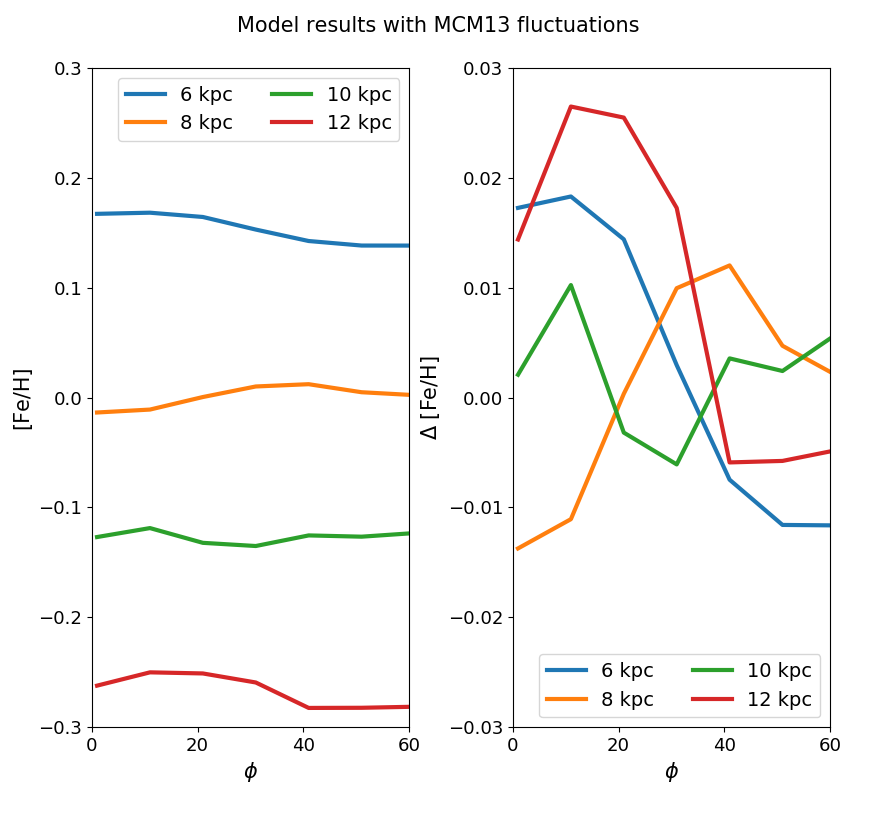}
   \caption{  {\it Upper Panel}: We present the average  Fe abundances of
     Galactic Cepheids presented by Genovali et al. (2014)  in bin of
     15$^{\circ}$  for the azimuthal coordinate $\phi$ at different
     Galactocentric distances.
      {\it Lower left Panel}: 
    Fe abundances as  functions of the azimuthal coordinates  computed
    at  6, 8, 10, 12 kpc predicted by  the chemical evolution model in
    which we implemented the 
    density fluctuation  by the MCM13 model.  {\it Lower right Panel}: residual of the
Fe abundances predicted by our model computed after subtracting the average radial gradient.
}
\label{CEP1}
\end{figure}

\section{Results}

In this section we apply our 2D model by using surface density fluctuations from the MCM13 chemo-dynamical model and from an analytical prescription.
\subsection{Density fluctuation from the MCM13 chemo-dynamical
  model}
In this section we present our results based on the new 2D chemical
evolution model including the density mass fluctuation extracted from the
chemo-dynamical model by MCM13.

Fig. \ref{mSFR}  shows the  galactic disc SFR  computed at 0.1, 0.7, 6, 11
Gyr    after the start of disc formation, for the
    chemical evolution  model in which we tested the effects of 
the density fluctuation by MCM13 in units of M$_{\odot}$ pc$^{-2}$ Gyr$^{-1}$. We 
notice that at early
times (i.e the ``1 Gyr'' case reported in the upper left panel), the
SFR is more concentrated in the inner Galactic regions,  the SFR in the innermost regions
decreases  and the outer parts become more star forming active  because of the ``inside-out'' prescription coupled with the
inclusion of the density fluctuation.
At  the Galactic epoch of 1 Gyr after  the start of disc formation,
regions with the same Galactocentric distances have
approximately the same SFR.
 Already after 0.7 Gyr of Galactic evolution, azimuthal star
formation  inhomogeneities are not negligible.  Concerning the panel
with  the model results  at 6 Gyr,
 azimuthal   inhomogeneities are evident, in particular at 8 kpc the
 ratio between the maximum and the minimum values assumed by the SFR is SFR$_{max}$ /SFR$_{min}$=6.72.

In Fig. \ref{mSFR}  the  bar and spiral arms features do not show up
clearly, especially in early times. This is caused by the adopted
inside out prescription (eq. \ref{tau}) which leads to huge differences between the
 SFRs computed in inner and outer regions.  
In Fig. \ref{1SFR},  the galactic disc  SFR($R$, $\phi$) is normalized to the maximum value
    SFR$_{R, \ max}$ of the annular region located at the Galactocentric distance
    $R$, i.e SFR($R$, $\phi$)/SFR$_{R,   max}$,
     computed at 0.1, 0.7, 6, 11
Gyr    after the start of disc formation, respectively.
Here, different features related to density perturbations originated
by spiral arms and bar can be noted.

In Fig. \ref{minchev_av} the main results related to the  present day oxygen
abundance azimuthal variation are presented.
The top panel shows  the azimuthal distribution of the residual of the
oxygen abundances computed with our chemical evolution model at 4, 8,
12, 16, and 18 kpc after subtracting the average radial gradient
(i.e. the
one obtained with the reference model without any density
perturbation).
Throughout this paper we adopt the photospheric values of Asplund
et al. (2009) as our solar reference.
We see that the behavior is in excellent   agreement with the
observations by S{\'a}nchez et al. (2015); indeed,  data show that outer regions 
display larger azimuthal variations, and the amplitude of the
risidual variations are of the order of 0.1  dex (see Figure 7 by
S{\'a}nchez et al. 2015) . In our model the maximum variations
are $\sim$ 0.12 dex for the chemical evolution models computed
at 18 kpc. Our results appears to have a bit less scatter.

 In the lower panel of Fig. \ref{minchev_av} we present our
``mock'' observations. We  draw oxygen abundances of different ISM regions
 at different Galactocentric distance at random  azimuthal
coordinates $\phi$.  Hence,  we add an error of $\sigma_{\phi}$=5$^{\circ}$ to
alleviate the fact that our model presents a  resolution of
10$^{\circ}$ in the azimuthal component $\phi$.
Moreover, the average observational  uncertainty associated to the  oxygen
abundances of $\sigma_{[O/H]}$ = 0.05 dex  provided by  S{\'a}nchez et al. (2015) has been considered.
 We define the ``new'' oxygen abundance  including these uncertainties as follows:
\begin{equation} 
 [\mbox{O/H}]_{new} =  [\mbox{O/H}]+  U([-\sigma_{[O/H]}, \sigma_{[O/H]}]),
 \label{Err1}
\end{equation}
where $U$ is the random generator function. 
Similarly, we implement the uncertainty in the azimuthal component through the following relation:
\begin{equation} 
 \phi_{new} = \phi +  U([-\sigma_{\phi}, \sigma_{\phi}]).
 \label{Err2}
\end{equation}

Here, it is clearly visible the similarity  between  the S{\'a}nchez et
al. (2015) observations
and our results. To summarize, the inclusions of density perturbations
taken from a self-consistent dynamical model at different Galactic
times, leads to  significant variations in chemical abundances in the
outer Galactic regions. 

In  Fig. \ref{minchev_grad} we show results for the present  day
abundance gradient (after 11 Gyr of evolution)  for  iron computed  for six azimuthal slices (as indicated) of width 10$^{\circ}$ at different
azimuthal coordinates. In the same plot is indicated with a shaded
grey area the maximum spread in the abundance ratio [Fe/H] obtained by
the azimuthal coordinates we considered (0$^{\circ}$, 60$^{\circ}$,
120$^{\circ}$, 180$^{\circ}$, 240$^{\circ}$, and 300$^{\circ}$). As a
consequence of the results presented above, the  shaded area is larger
towards external regions.
We also overplot the data from Genovali et al. (2014) in order to compare to our model predictions. 
We  notice that the  predicted gradient is slightly steeper than the
observed one in the external Galactic regions.
However,   we notice that  the model
lines pass within  in the data standard deviation computed  dividing the data by  Genovali et al. (2014) in six radial bins.

In Fig. \ref{minchev_grad_ev}  we tested the effects of
chemo-dynamical fluctuations on the time evolution of the oxygen abundance gradient at a fixed azimuth ($\phi$=0$^{\circ}$).
In agreement with Minchev et al. (2018) the abundance gradient flattens  in time, because of the chemical evolution
model assumptions. 
As shown by Spitoni et al. (2015) and Grisoni et al. (2018), the
inclusion of radial gas flows can in lead to even steeper gradients in
time during the whole Galactic history.

\begin{figure}
  \includegraphics[scale=0.43]{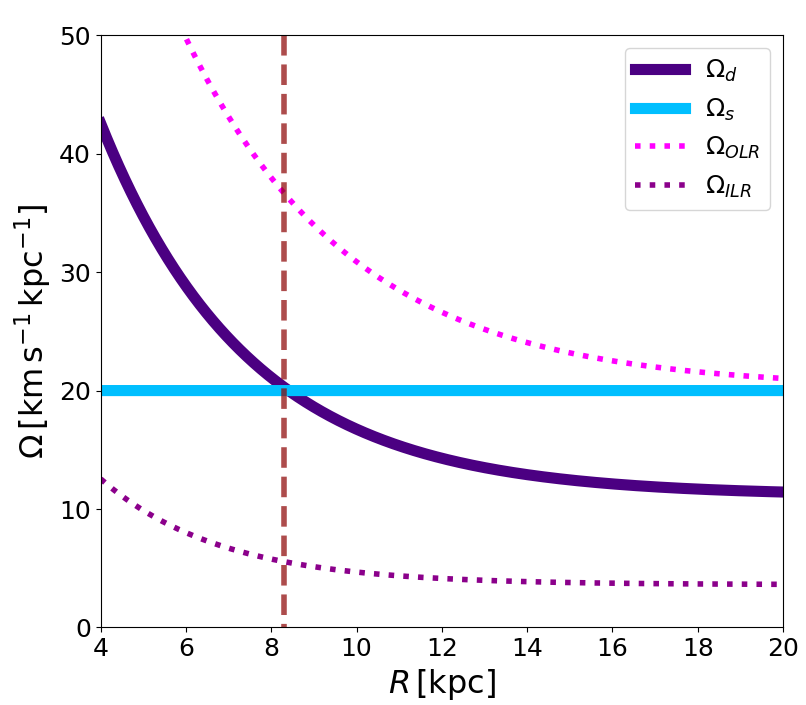}
   \caption{ Spiral pattern speed
$\Omega_s$  and disk  angular velocity
$\Omega_d$ computed  by Roca-F{\`a}brega  et al. (2014) are indicated with light blue and violette lines,
respectively. With the vertical long dashed red line we show the position of the
corotation radius located at the Galactocentric distance
$R=8.31$ kpc. Outer and Inner Lindblad resonances extracted by
Roca-F{\`a}brega  et al. (2014) simulation are also drawn with dotted
magenta and dotted purple lines, respectively. }
\label{omega}
\end{figure}

\begin{table}
\caption{  Different spiral arm models tested with our 2D chemical
  evolution model for the disc changing the number of spiral arms $m$
  (second column), the the pitch angle $\alpha$ (third column), and
  finally the spiral pattern speed  $\Omega_{s}$  is indicated in the
    last column.}
 \label{TMW}
\begin{center}
\begin{tabular}{c|cccc}
  \hline
\\
 Models &$m$ &$\alpha$&  $\Omega_{s}$\\
&  & & [km s$^{-1}$ kpc$^{-1}$]\\  
  
\hline
\\
S2A & 2  & 15$^{\circ}$ & 20 \\
S2B & 2  & 15$^{\circ}$ & 17.5\\
S2C & 2  & 15$^{\circ}$ & 15\\
S2D & 2  & 15$^{\circ}$ & 13.75\\
S2E & 2  & 15$^{\circ}$ & 12.5\\
S2F& 2  & 15$^{\circ}$ & 25\\
S2G & 2  & 7$^{\circ}$ & 20 \\
S2H & 2  & 30$^{\circ}$ &  20\\
S1A & 1  & 15$^{\circ}$ & 20 \\
S1B & 1  & 15$^{\circ}$ & 17.5\\
S1C &1  & 15$^{\circ}$ & 15\\
S1D & 1  & 15$^{\circ}$ & 13.75\\
S1E & 1  & 15$^{\circ}$ & 12.5\\
S1F& 1  & 15$^{\circ}$ & 25\\

\hline

 \hline
\end{tabular}
\end{center}
\end{table}

\begin{figure}
 \includegraphics[scale=0.44]{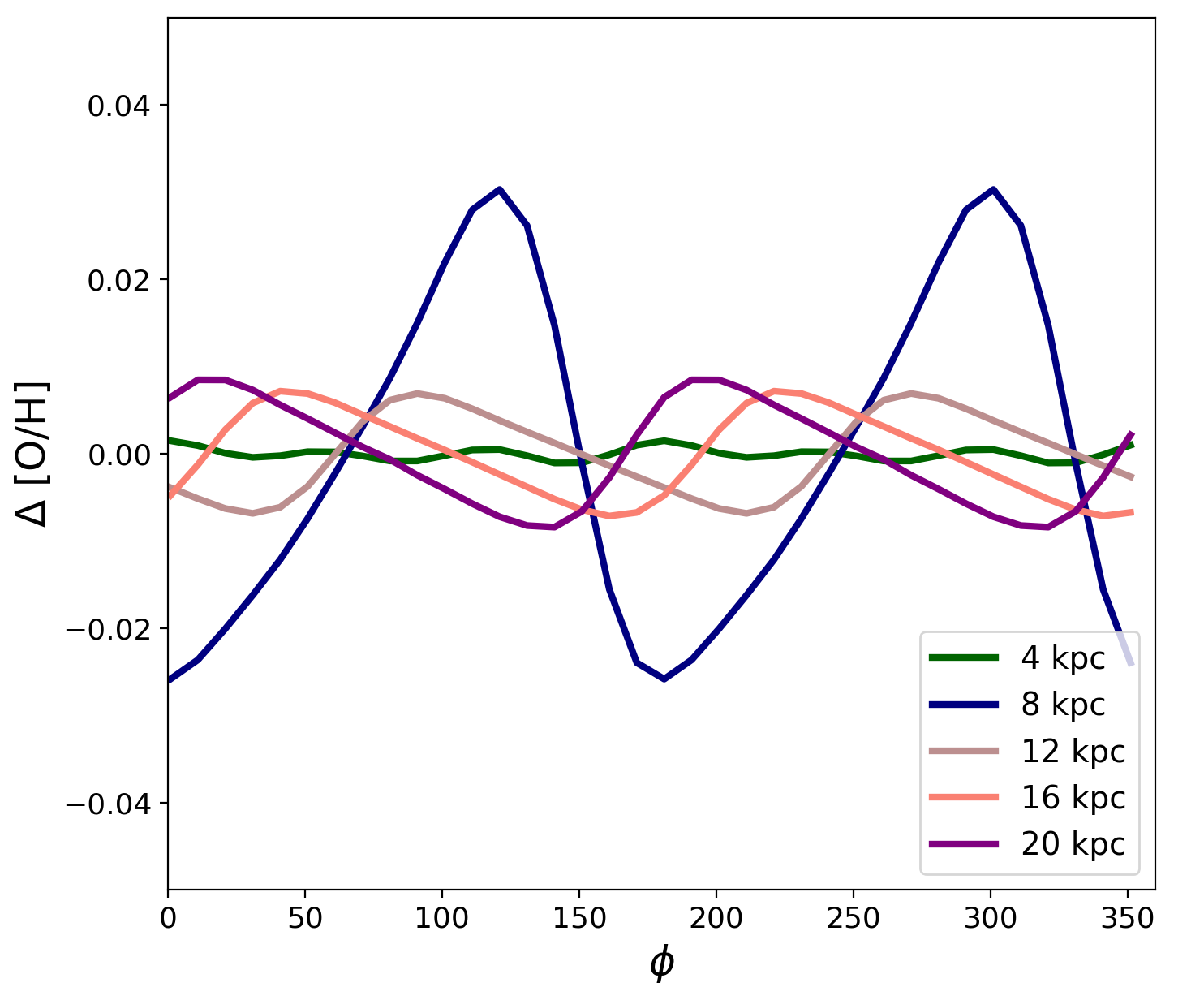}
  \includegraphics[scale=0.44]{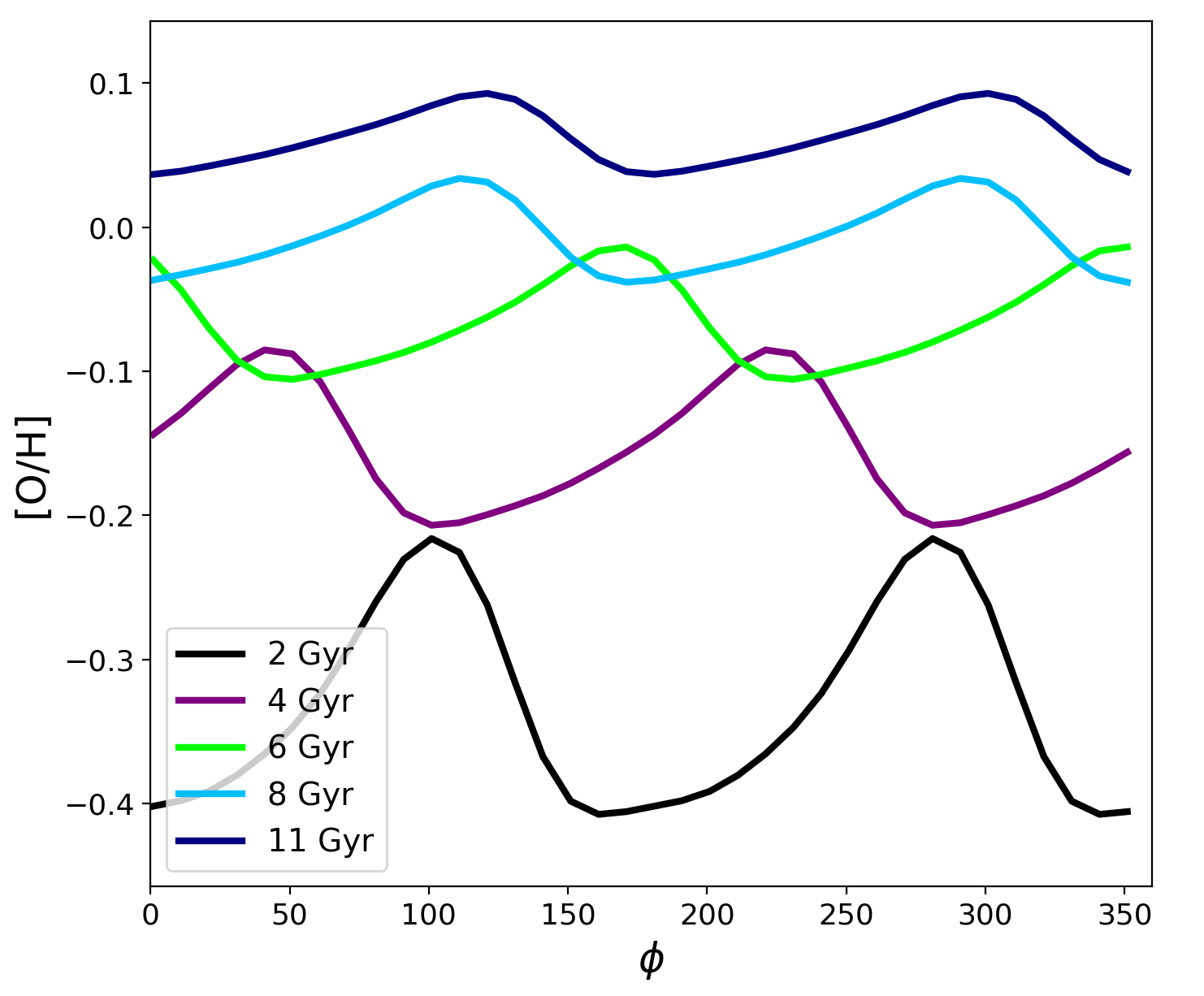}
   \caption{  
Results for the chemical evolution model in which we consider
the density fluctuation associated with the analytical spiral arm
formulation. {\it Upper Panel}: The azimuthal distribution of the residual of the
oxygen abundances computed with our chemical evolution model at 4, 8,
12, 16, and 20 kpc (after subtracting the average radial gradient for
a model with $R_S$=7, $R_D$=3.5, $\Sigma_0$=20, $\nu$=1.1, 
 $\Omega_s$=20  km
s$^{-1}$ kpc$^{-1}$, and  $m$=2 spiral arm (model S2A in Table 1).  {\it Lower Panel}: the
time evolution of the   [O/H] abundance as  a function of the
azimuthal coordinate computed at  8 kpc. 
}
\label{SA}
\end{figure}
\begin{figure}
 \includegraphics[scale=0.48]{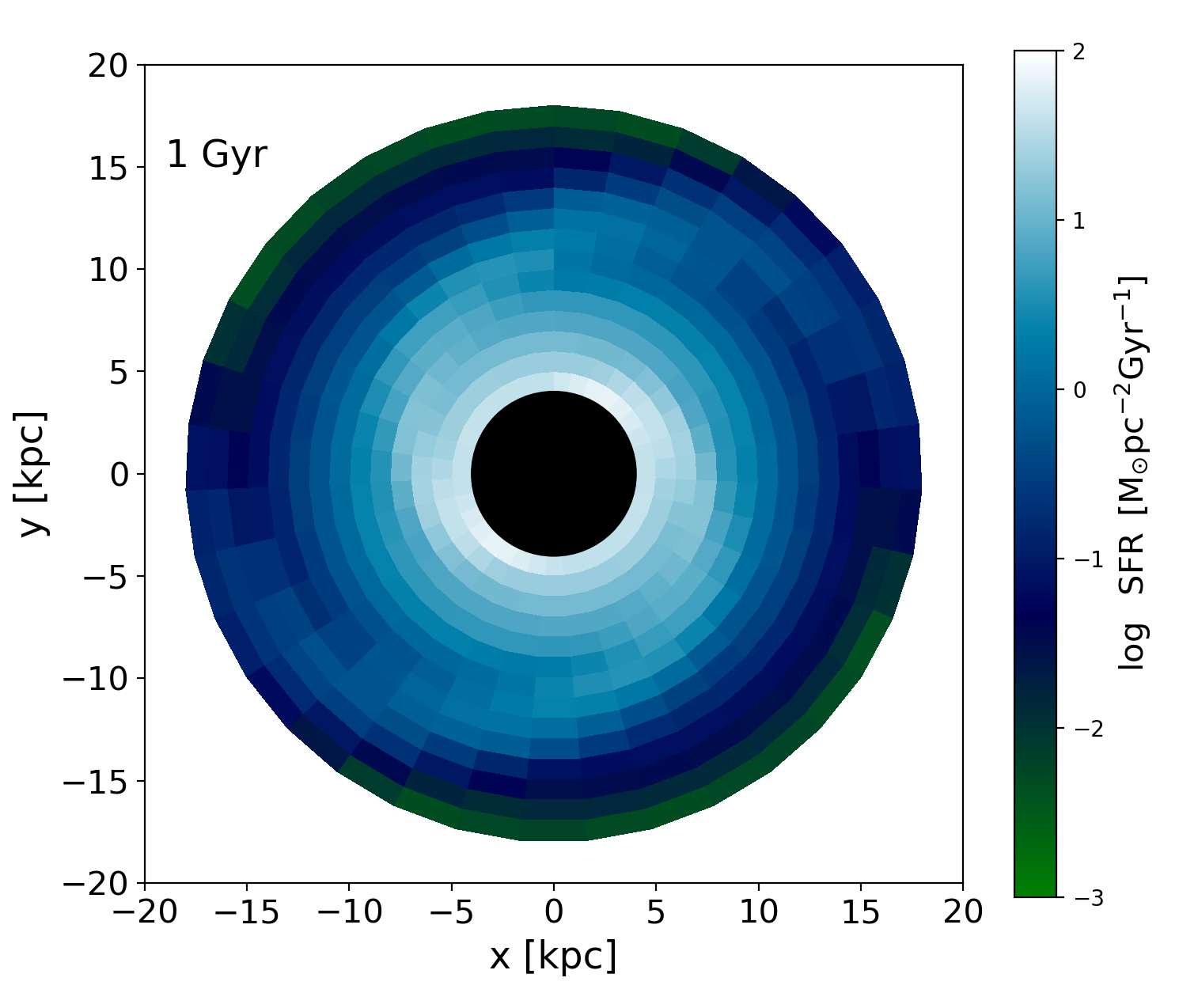}
 \includegraphics[scale=0.48]{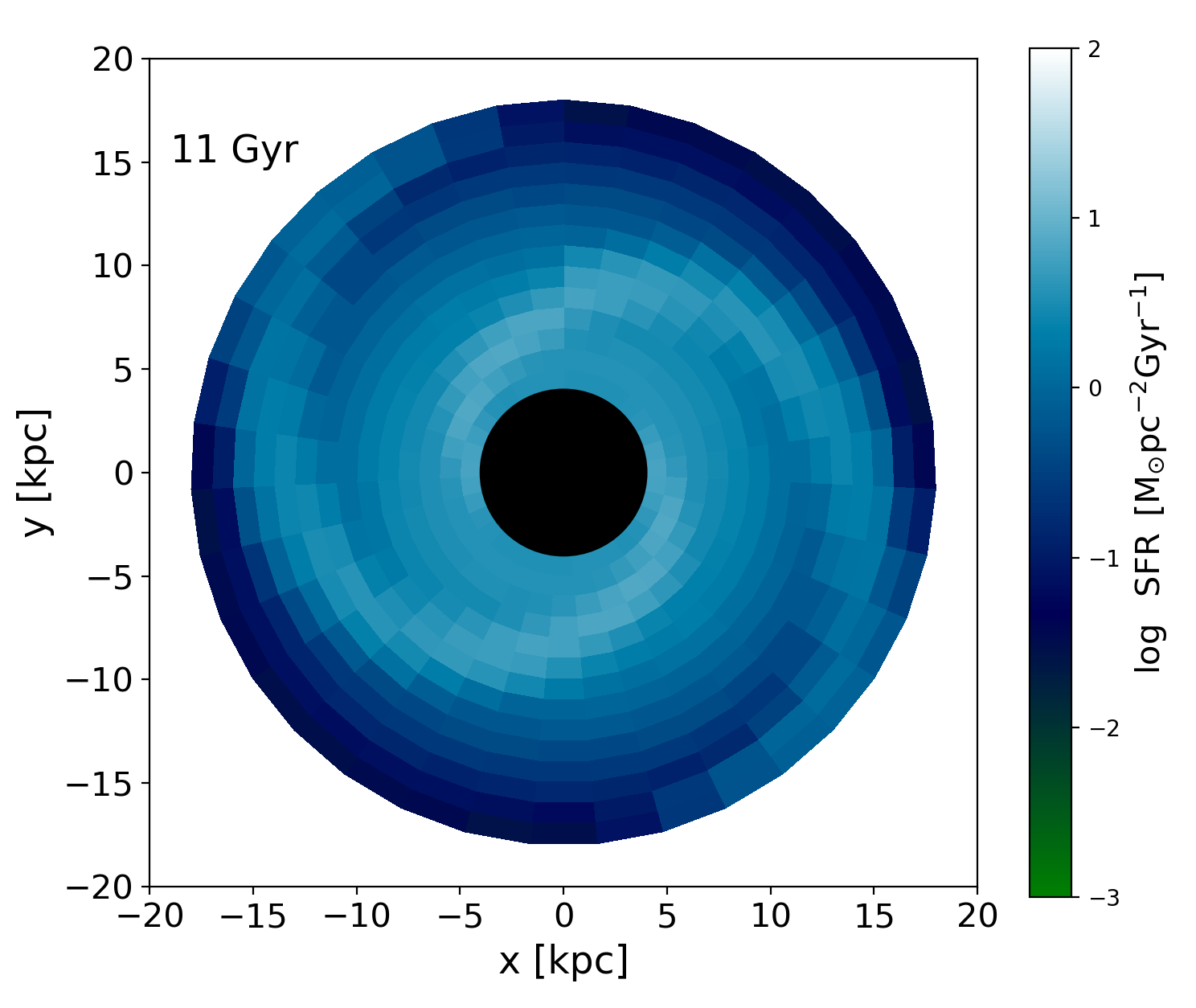}
  \caption{  {\it Upper Panel}:  Galactic disc SFR resulting from model S2A after 1 Gyr of evolution (see Table 1 and text for model details). The color code indicates the SFR in units of     M$_{\odot}$ pc$^{-2}$ Gyr$^{-1}$.   
{\it Lower Panel}: same but computed at 11 Gyr.}
\label{SASFR}
\end{figure}

\begin{figure}
\includegraphics[scale=0.40]{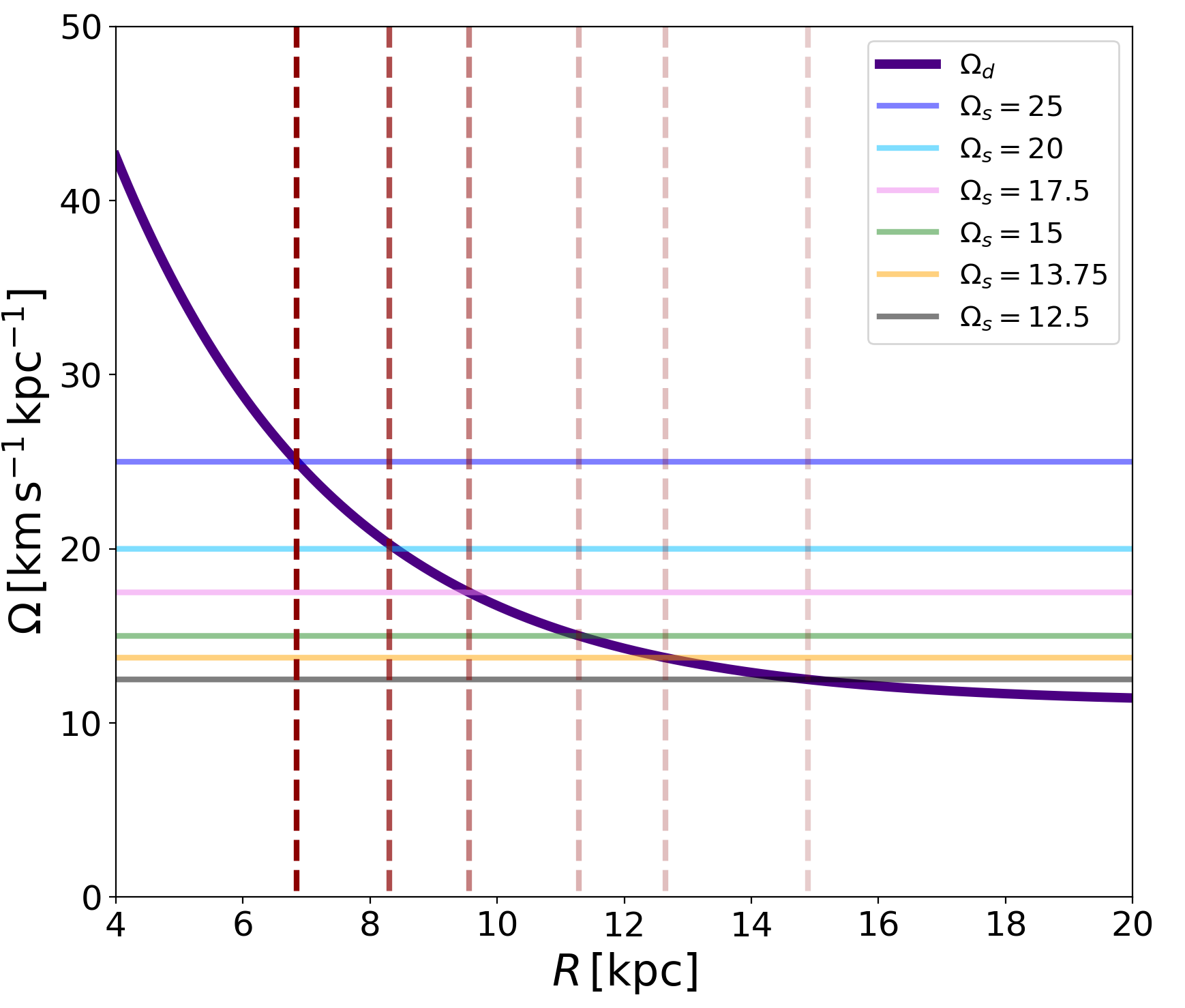}
   \caption{ 
 Disk  angular velocity
$\Omega_d$ computed  by Roca-F{\`a}brega  et al. (2014) is indicated
with light blue and violette lines. With different horizontal solid
lines are indicated the spiral pattern speed $\Omega_s$ adopted in our
models (see text and Table 1 for model details). The vertical long dashed  lines  show the positions of the
corotation radii assuming different $\Omega_s$ values.}
\label{DOS}
\end{figure} 

\begin{figure*}
\centering
\includegraphics[scale=0.5]{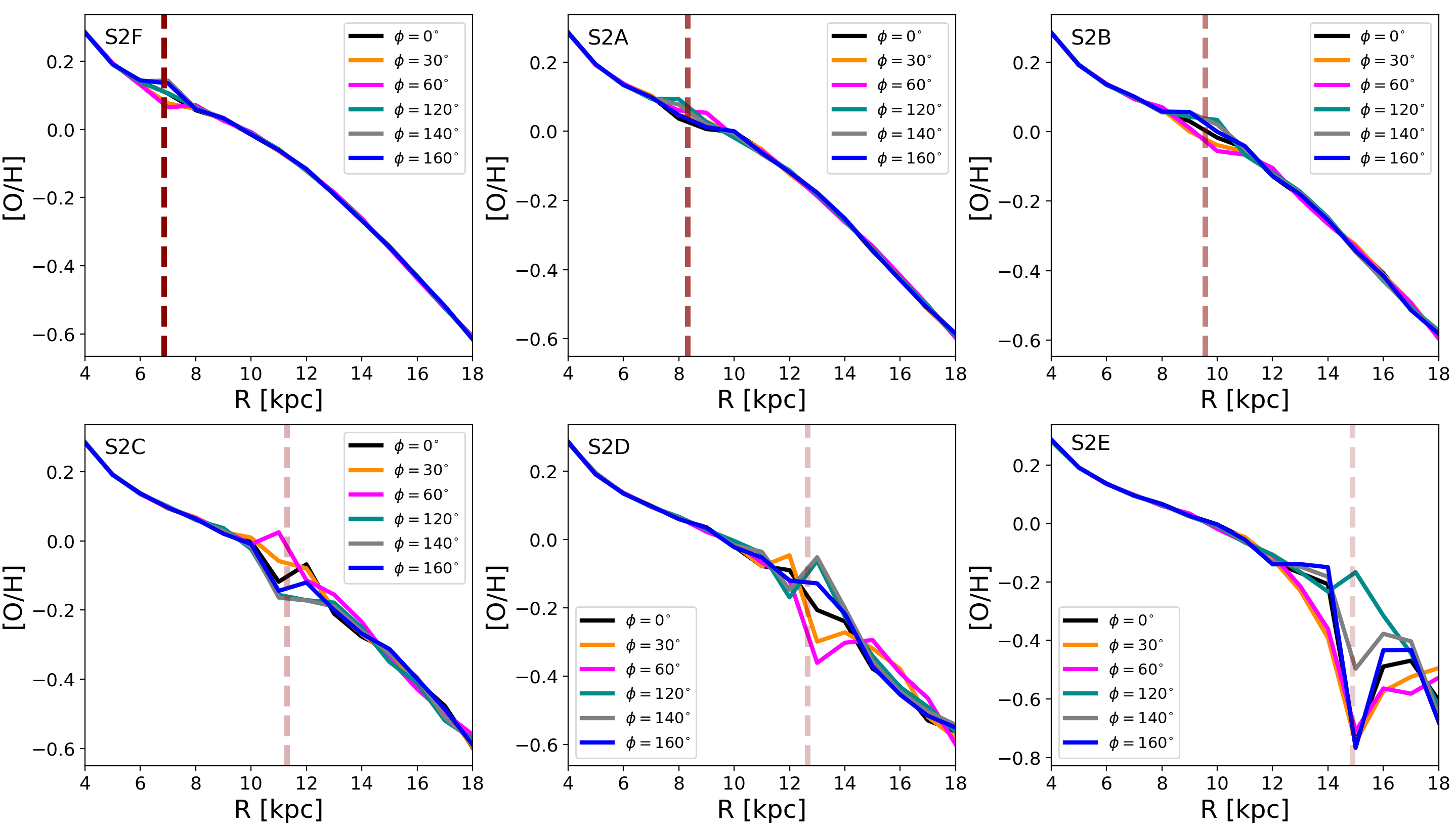}
 \caption{ The present day oxygen abundance gradients for different azimuths,
predicted by chemical models with  spiral multiplicity $m = 2$ and different
spiral pattern speed $\Omega_{s}$ (see Table 1 for model details).
In each panel the dashed vertical line indicates the location of the
corotation radius. It is clear that allowing for multiple spiral patterns propagating through the disk at the same time will affect the entire disk, similarly to the case of the MCM13 model.}
\label{S2_grad}

\end{figure*}

\begin{figure}
\includegraphics[scale=0.5]{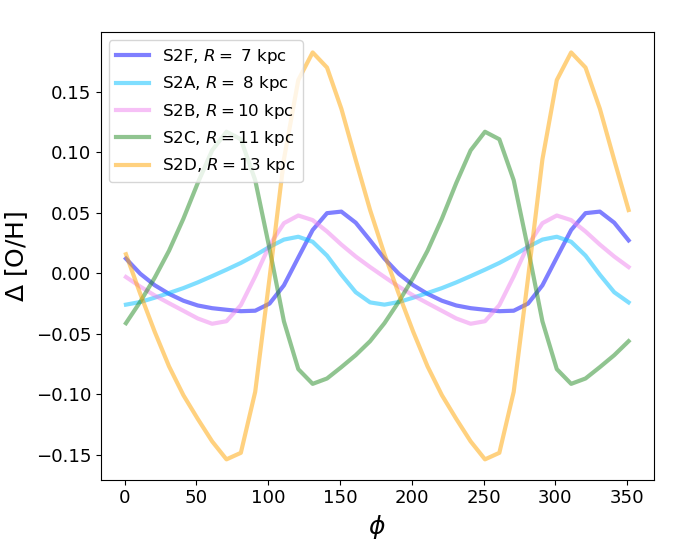}
   \caption{ 
Present day residual azimuthal variations in oxygen abundance for the corotation regions (as indicated) of the different pattern speeds shown in Fig. \ref{S2_grad}. An increase in the effect is found as the corotation shifts to larger radius, i.e., for slower spiral patterns. Such a set of spirals with progressively slower patterns speeds as radius increases, can be a realistic representation of a galactic disk. 
}
\label{CS2}
\end{figure}

\begin{figure}
   \includegraphics[scale=0.47]{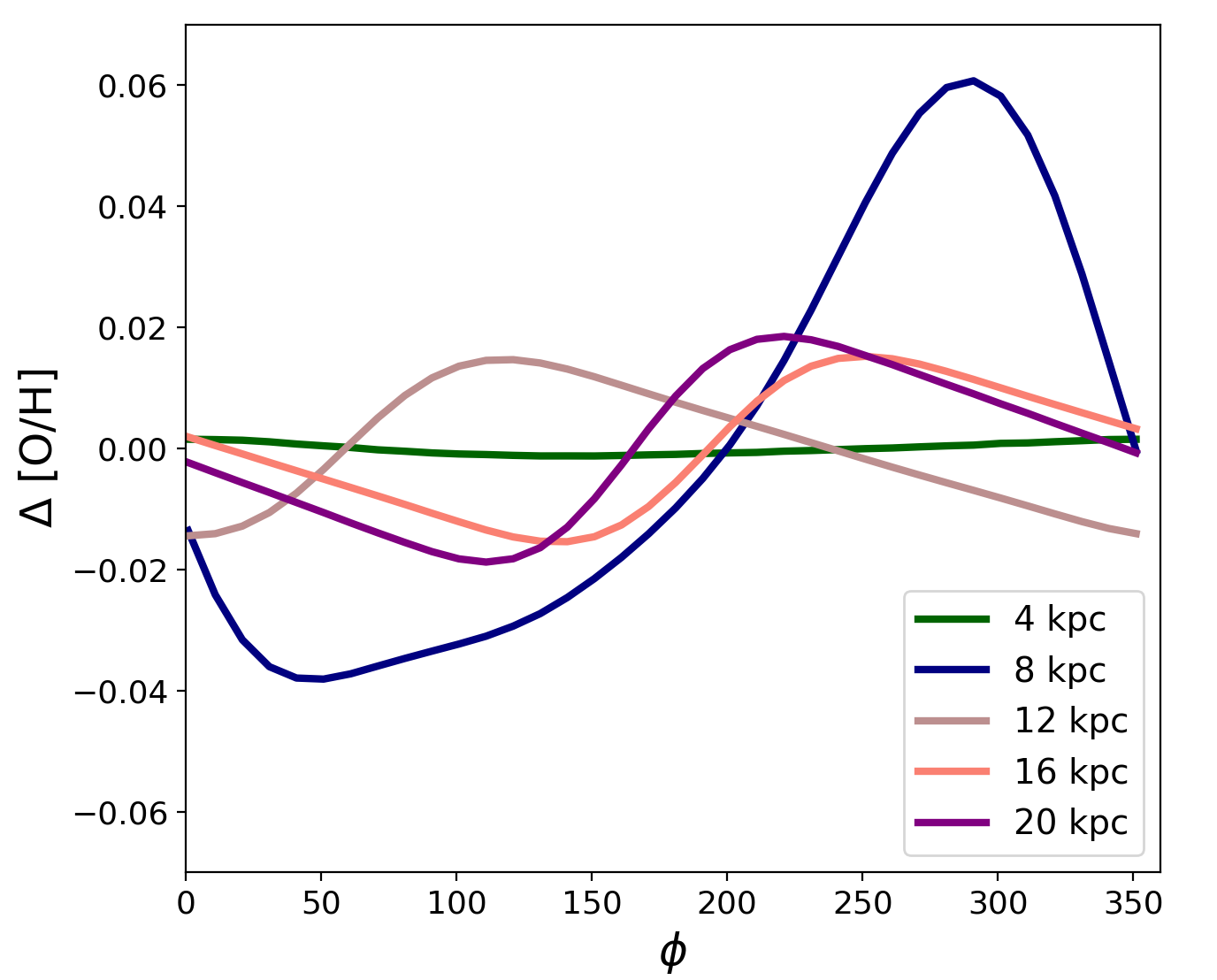}
\includegraphics[scale=0.47]{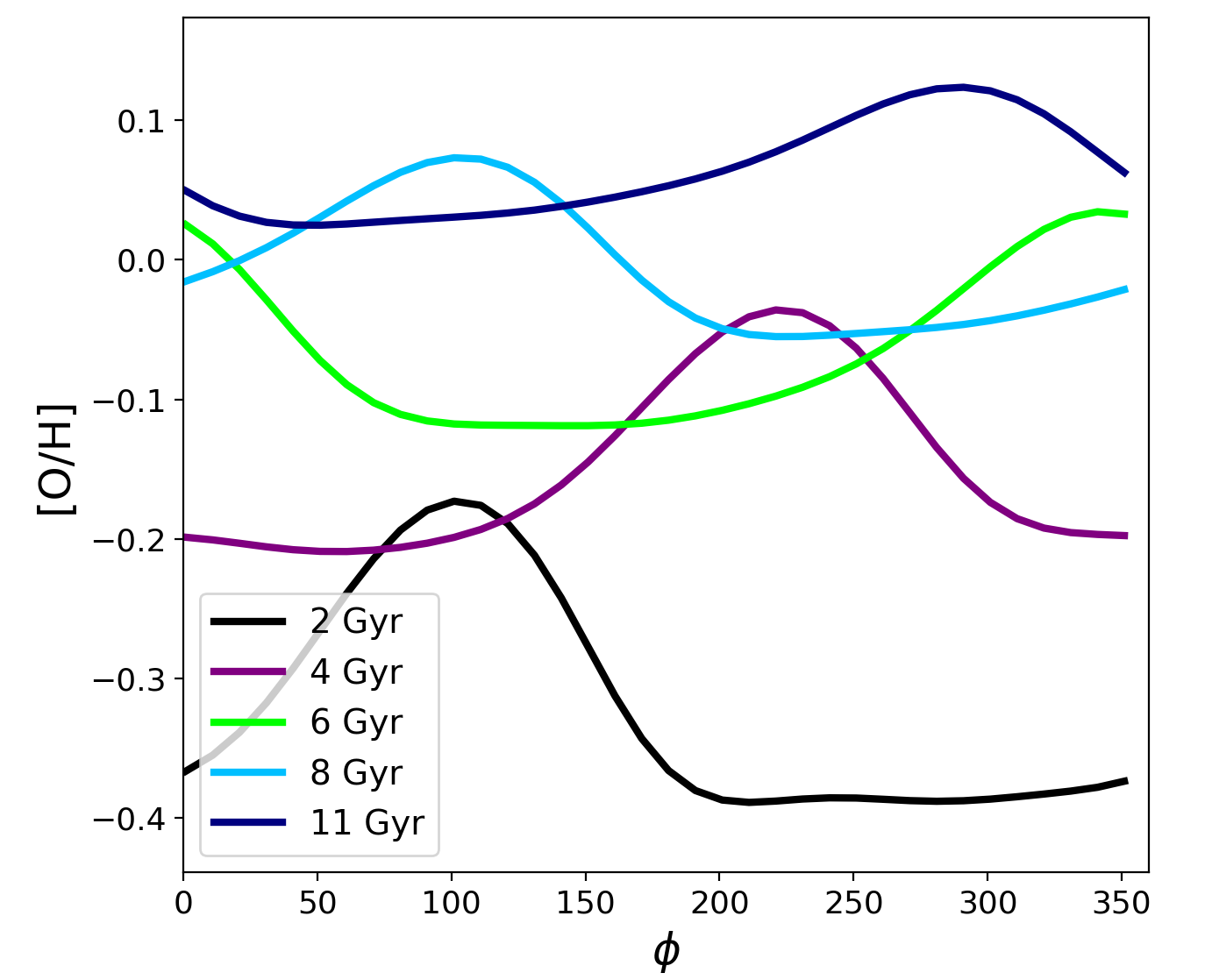}
 \caption{ As in Fig. \ref{SA}  but for model S1A, with with multiplicity
$m$=1 of spiral arms.}
 \label{S1A}
\end{figure}

In Fig. \ref{CEP1} we compare the average iron abundance azimuthal
variation in bins of $\phi$=15$^{\circ}$  presented by Genovali et al. (2014)  computed at 6, 8, 10,
and 12 kpc, respectively  with our 2D chemical evolution model, resulting from the MCM13 density variations.
We see that the observed azimuthal variations are for limited
Galactocentric distances (6-12 kpc) and with a narrow range of azimuthal coordinates.
Although it is evident that the observed amplitude of azimuthal variations are larger than the ones 
predicted by our models,  more precise Galactic
Cepheid data are required to  make firm conclusions. 

Moreover,  other dynamical
processes that we have not considered in this work had maybe played important
roles in the evolution and in the building up  of the Galactic
gradients and their azimuthal variations - 
radial migration processes can already introduce some variations in about a Gyr (Quillen et al. 2018).

\begin{figure*}
\centering
\includegraphics[scale=0.5]{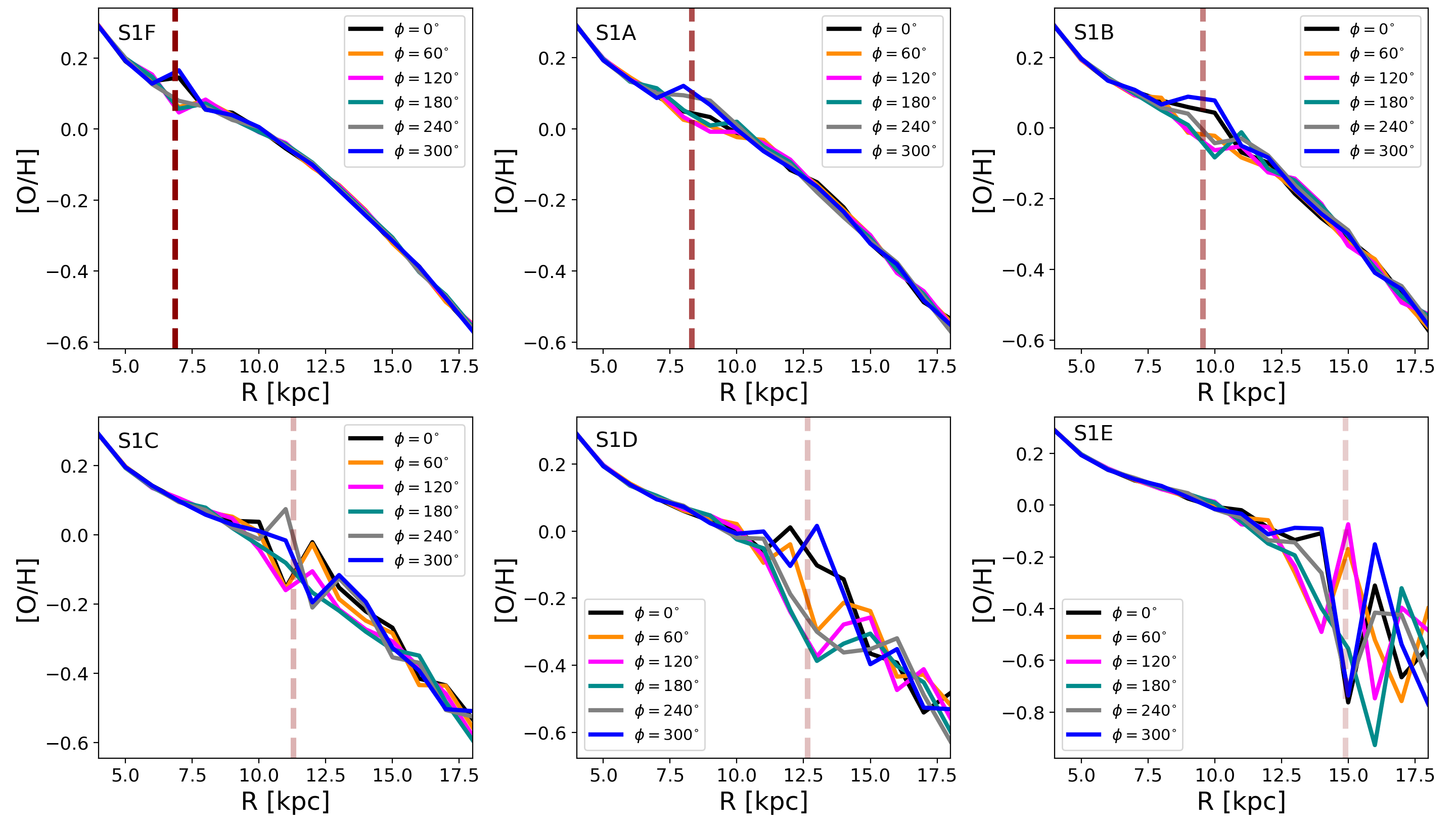}
    \caption{ 
As Fig. \ref{S2_grad}, but for an $m$=1 spiral.
}
\label{S1_grad}
\end{figure*}

\begin{figure}
 \includegraphics[scale=0.5]{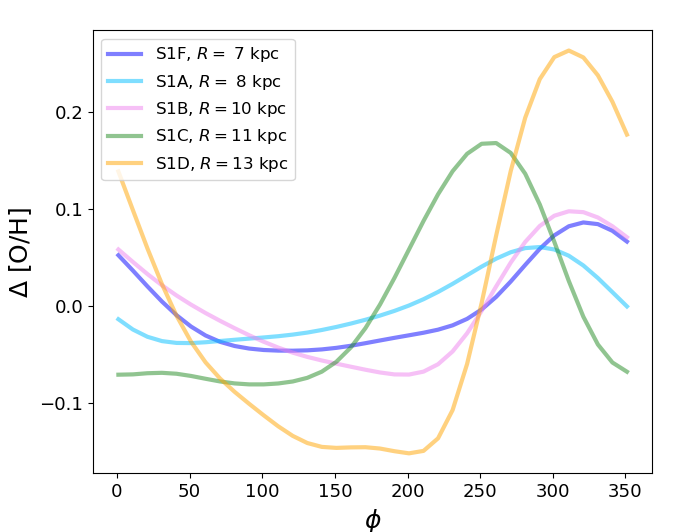}
    \caption{ 
 As in Fig. \ref{CS2} but for 
  models with $m=1$ multiplicity (see Table 1).}
\label{CS1}
\end{figure}

\subsection{Density fluctuations from an analytical spiral arm formulation}
In this Section
we discuss 
the results of  chemical evolution models with  only analytical
prescriptions for spiral arm density perturbations without including
any density fluctuations from chemo-dynamical models.
The primary purpose here is   to test the effect of regular
perturbations  (i.e. spiral arms evolution described by an analytical formulation) on the chemical
evolution of a Milky Way like galaxy.
We underline that the results showed  in the previous Section reflect more closely the complex behavior of the Milky Way.
However, we are also interested
to explore different
spiral arm configurations which could 
characterize external galactic systems by varying  the  free parameters of the analytical
expression of the spiral arms.
In particular, we will show the effects on the
azimuthal variations of abundance gradients for  oxygen  by varying:
\begin{enumerate}[i)]
\item  the multiplicity $m$  of  spiral arms;
   \item
 the spiral pattern speed, $\Omega_s$;
  \item the pitch angle $\alpha$.

\end{enumerate}

For all model results that  will be presented we 
assume  the following Cox \& Gomez (2002) prescriptions:  the radial scale length of the drop-off in density amplitude of the arms fixed at the value of  $R_S=7$ kpc,
the pitch angle is assumed constant at
$\alpha=\ang{15}$, and  the surface arm
density $\Sigma_0$ is 20 M$_{\odot}$ pc$^{-2}$ at the fiducial
radius $R_0=8$ kpc; finally we assume $\phi_p(R_0)=0$.

The disk rotational velocity $\Omega_d(R)$ has been extracted from the simulation
by  Roca-F{\`a}brega et al. (2014)  (see their left panel of Figure
1).  The exponential fit of  $\Omega_d(R)$ variations as a
function of the Galactocentric  distance $R$ (expressed in kpc) is:
\begin{equation}
\Omega_d(R)=98.93 \, e^{-0.29 \, R}+ 11.11 \mbox{  [km } \mbox{s }
^{-1}\mbox{kpc}^{-1}]. 
\label{eqom}
\end{equation}
We start by adopting the constant pattern angular velocity   $\Omega_s$ =
20 km s$^{-1}$ kpc$^{-1}$ consistent with the  Roca-F{\`a}brega et
al. (2014) model. Similar value was first estimated from
moving groups in the U-V plane by Quillen \& Minchev (2005, 18.1
$\pm$ 0.8 km s$^{-1}$ kpc$^{-1}$)   and a summary of derived values for the Milky Way can be found in Bland-Hawthorn \&
Gerhard (2016).  In Fig. \ref{omega} we show  the $\Omega_s$
and $\Omega_d(R)$ quantities  as well as the Outer and Inner Lindblad 
resonances as a function of the Galactocentric
distance,   the corotation radius is located
at 8.31 kpc.

\subsubsection{Results with a single analytical spiral pattern}
 We begin our analysis discussing the results obtained with model S2A (see Table 1), which has a pattern speed of $\Omega_s$ =
20 km s$^{-1}$ kpc$^{-1}$, placing the corotation resonance at the solar radius.

 The upper panel of Fig.  \ref{SA}  shows the 
the oxygen abundance residual azimuthal variations after 11 Gyr of disc evolution for different Galactocentric distances. The average radial gradient is subtracted. 
As expected, larger abundance azimuthal variations are found near  the corotation radius.
In  this region the chemical enrichment should be more
efficient  due to the
lack of the relative gas-spiral motions. Higher SFR at the
corotation radius  caused by locally higher gas overdensity lasts for a longer time,    therefore more
massive stars can be created and more metals can be ejected into the
local ISM under the spiral arm passage.

At 8 kpc we have  $\Delta$[O/H]
  $\approx$ 0.05 dex. For  other Galactocentric distances, away from the corotation,
  variations are much smaller. In the lower panel of
Fig. \ref {SA} we present the temporal evolution of the oxygen
abundance azimuthal variations for the model S2A as a function of the
azimuthal coordinate $\phi$ computed  at 8 kpc.
As expected, larger inhomogeneities are present at early times, decreasing in time.

 As discussed
in Section 2.2, we  assume that during the
Galactic evolution the ratio between the amplitude of the spiral
density perturbation and the total surface density computed at the same radius $R$, remains
constant in time. However, this analytical approach is not capable to
put constraints on the temporal evolution of pattern speed.

 Galactic chemical evolution is an integral process in time. The stronger spiral structure induced azimuthal variations at early times are, therefore, washed out by phase mixing.

 Fig. \ref{SASFR}  depicts the SFR 
  after 1 Gyr of evolution (upper panel) and  at the present time (lower panel)   on the
galactic plane computed with the model S2A. Here, it is evident the way
in which 
the spiral arm density perturbation affects and modulates  SFR computed at the present time in unit of
M$_{\odot}$  pc$^{-2}$   Gyr$^{-1}$. The shape of the two spiral arm over-densities is
clearly visible in the SFR.  This is in contrast to our results using
the MCM13 density fluctuations (see Fig. \ref{mSFR}), where multiple
spiral density waves were present. Moreover, we can  appreciate the inside-out disk formation:
at later times the external regions become star formation active.

 \subsubsection{The effect of   different pattern speeds}

In this Section we vary the spiral pattern speed,
which has the effect of shifting the corotation resonance in
radius. We argue that a combination of multiple spiral modes with
different pattern speeds can be a realistic representation of a
galactic disk.
The horizontal and vertical lines in Fig. \ref{DOS} show the different pattern speeds and corresponding corotation radii, respectively, used in this Section:  it is clear that smaller $\Omega_{s}$ values lead
to a more  external corotation radius.

In Fig.   \ref{S2_grad}  we show the 
oxygen abundance gradients computed at 
different azimuths after 11 Gyr of disk evolution for models with spiral multiplicity
$m$ = 2 and different spiral pattern speed $\Omega_{s}$  (see Table 1
for model details).

We notice that the more the corotation radius is shifted towards the
external Galactic regions the more the  oxygen azimuthal abundance
variations are amplified  near the corotation radius.
This result is reasonable in the light of our previous findings
presented above with our model assuming chemo-dynamical fluctuations by
MCM13. We recall that larger variations in the chemical
abundance of  outer galactic regions have been found by observations in external galaxies (S{\'a}nchez et
al. 2015).

In Fig. \ref{CS2} we show the present day azimuthal residual of the
oxygen abundances after subtracting the average radial gradient
computed for the Galactic annular regions which include the relative
corotation radius for the following models with $m=2$ multiplicity:
S2A, S2B, S2C, S2D, S2F (see Table 1 for other parameter details).
The model S2D computed at $R=$13 kpc has $\Delta$[O/H]$\approx$ 0.32
dex. Already in regions not so far from the solar neighbourd, the
variations are important, i.e., model S2C whose corotation resides in
the annular region centered at $R=$ 11 kpc, presents an oxygen
abundance variation of $\Delta$[O/H] is $\approx$ 0.20 dex.  

As
discussed in Setion 2.3, it is well accepted that multiple patterns can
be present in galactic disks (e.g., Meidt et al. 2009) including our own
Milky Way (Minchev \&
Quillen 2006, Quillen et al. 2011), with slower patterns shifted to outer radii. This will have the effect of
placing the corotation regions very similarly to what Fig. \ref{CS2}
presents and having corotating arms at all radii as found by Grand et al. (2012), Hunt et al. (2019). Therefore, the increasing scatter in abundance with galactic
radius can be explained as the effect of multiple patterns propagating
at the same time. Note that radial migration will introduce additional
scatter, that can in principle be accounted for.

\begin{figure}
\includegraphics[scale=0.5]{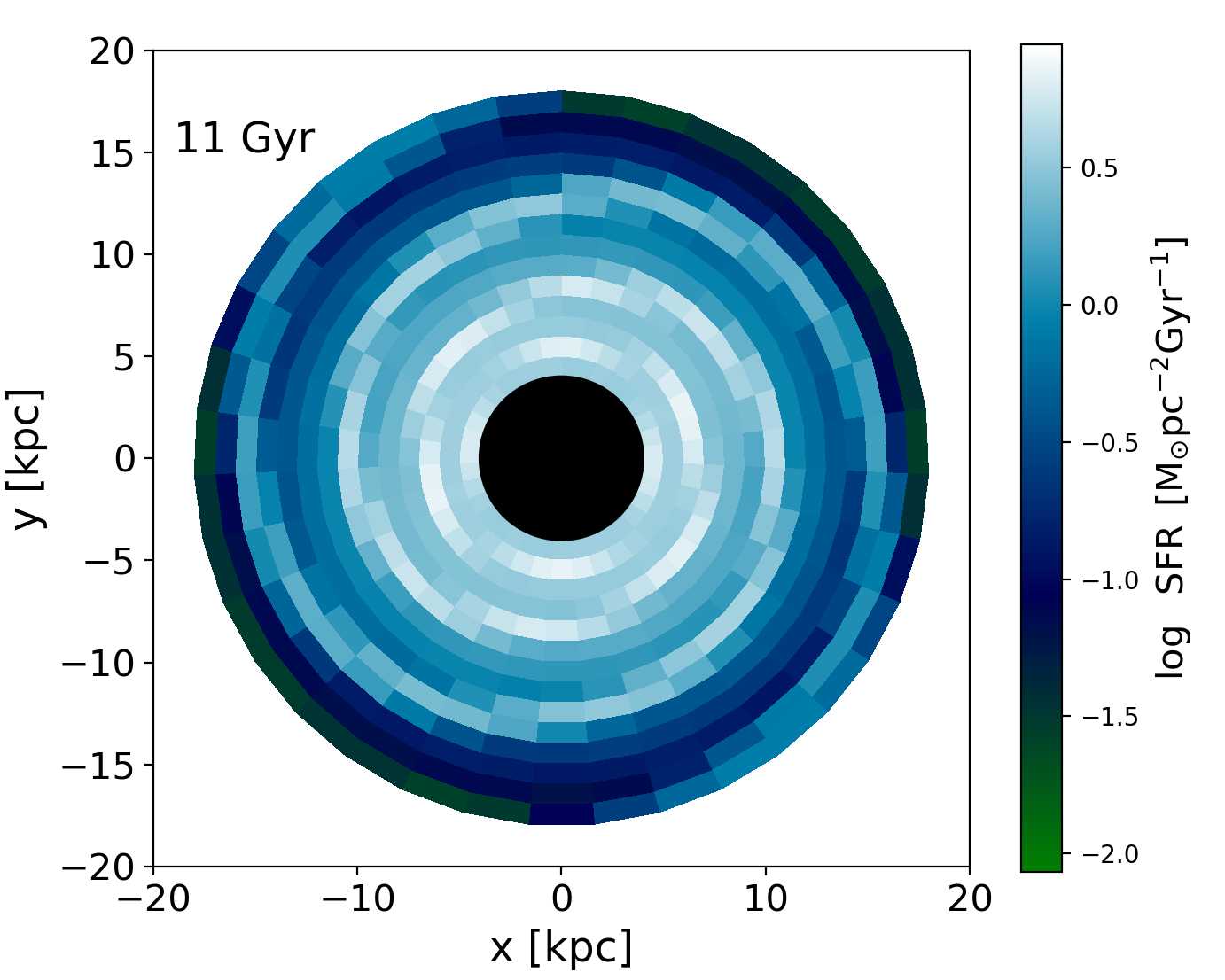}
  \includegraphics[scale=0.5]{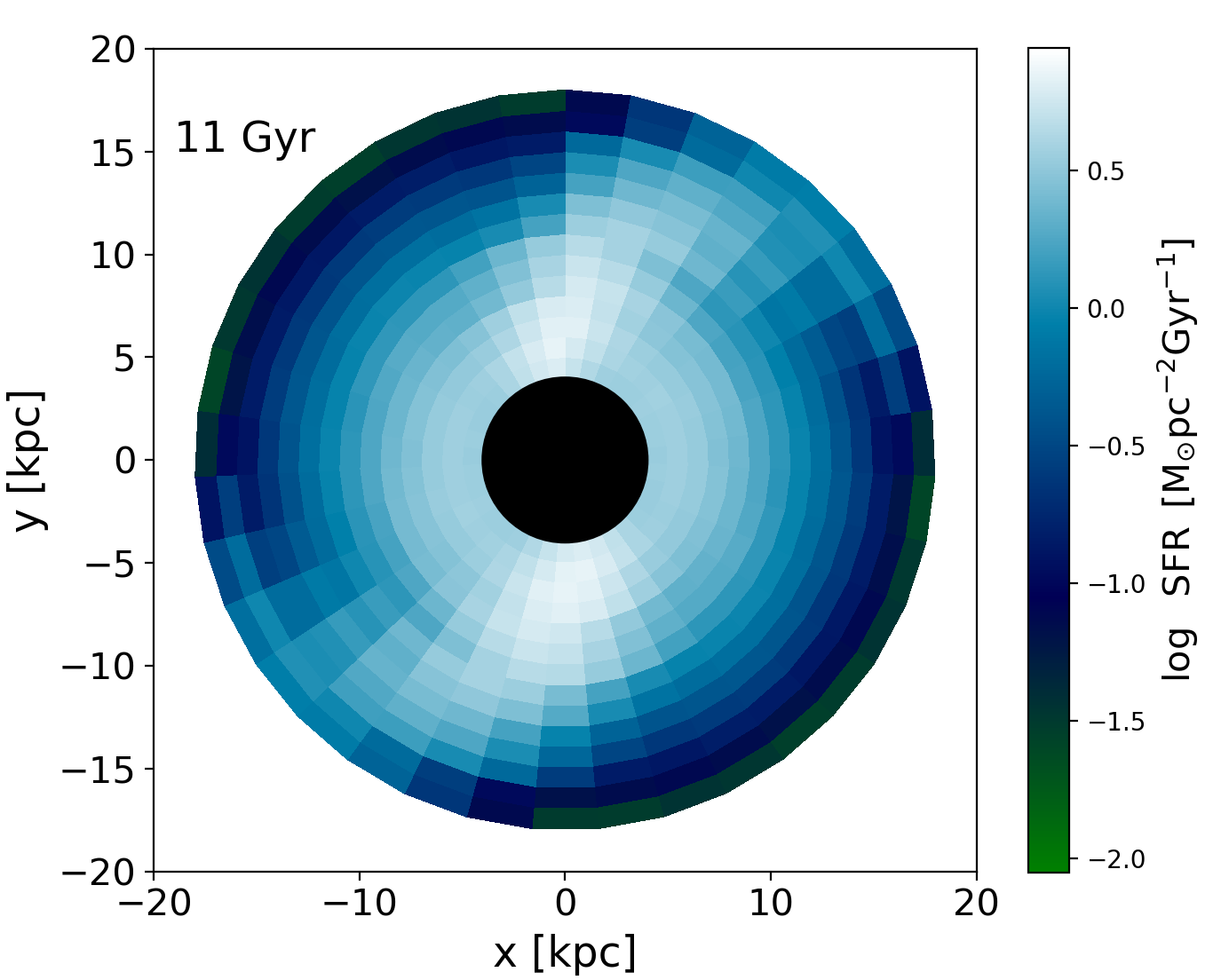}
\caption{  {\it Upper Panel}: The  Galactic disc SFR related to the
    model S2G computed after 11 Gyr of Galactic evolution
    (see Table 1 and text for model details)  with a pitch
    angle $\alpha=7^{\circ} $. 
The color code indicates the SFR  in  units of M$_{\odot}$ pc$^{-2}$ Gyr$^{-1}$.   
{\it Lower Panel}: as the upper panel but for the model S2H where the
pitch angle $\alpha$ is 30$^{\circ}$.}
\label{SESFSFR}
\end{figure}
\begin{figure}
\includegraphics[scale=0.47]{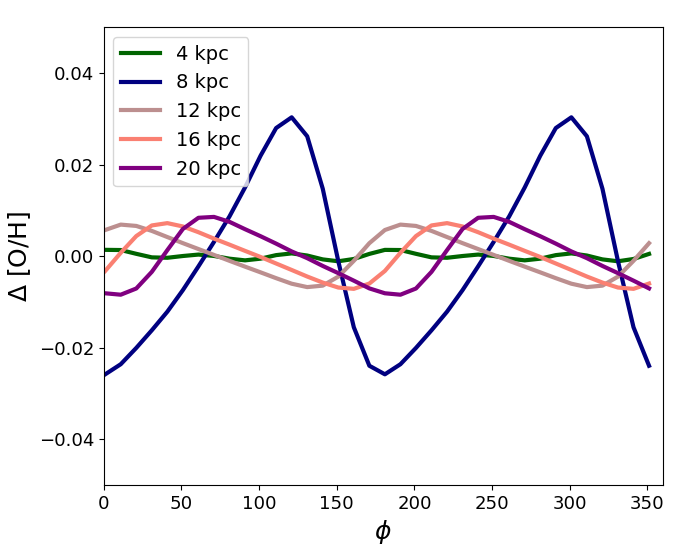}
  \includegraphics[scale=0.47]{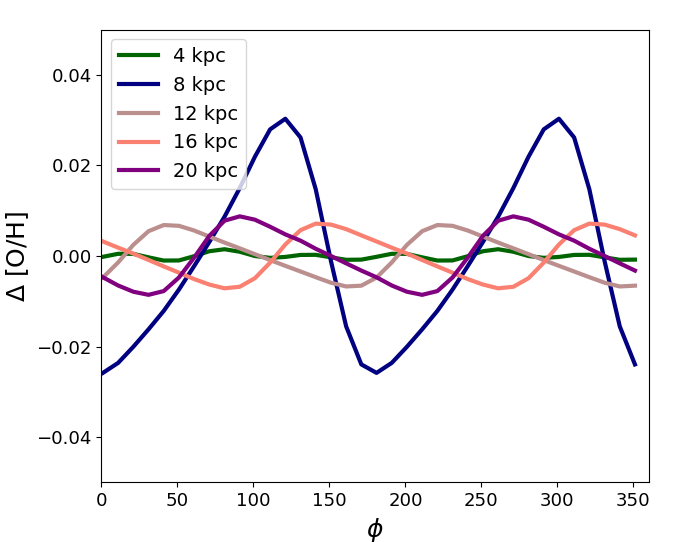}
   \caption{ Effects of different pitch angles $\alpha$ on the azimuthal distribution of the residual of the
oxygen abundances computed with our chemical evolution model at 4, 8,
12, 16, and 20 kpc In the upper panel the
pitch able is set at the value of 7$^{\circ}$ (SE model in Table 1), while in the lower
panel $\alpha=30 ^{\circ}$ (SF model in Table 1). }
\label{SE_SF}
\end{figure}

\subsubsection{Results with an $m$=1 spiral pattern}

We want to test whether   the intensity of the amplitude of
 the azimuthal chemical
abundance variations is dependant on the number $m$ of spiral
arms. In Table 1 we label as model  S1A a model identical to the model
 S2A
 but with an $m$=1 spiral structure, i.e., having only one spiral
 arm. Such a mode arises naturally  from the coupling of $m$=2 and
   $m$=3 modes as found by Quillen et al. (2011) and Minchev et al. (2012a) using pure N-body and SPH simulations,  and is seen in external galaxies (Zaritsky \& Rix 1997).

In  the upper panel of Fig. \ref{S1A} we notice that the abundance
variations are  larger than the ones obtained with the same  model
but $m$=2 (upper panel of Fig.   \ref{SA}): 
a fluctuation of about $\Delta$[O/H]=0.1 dex is seen at the corotation
radius ($\sim$ 8 kpc).

In the same Figure  is presented the time evolution of  azimuthal
abundance   inhomogeneities   for oxygen computed at 8
  kpc with the  model S1A at 2, 4, 6, 8, and 11 Gyr.

In Fig.   \ref{S1_grad}  we have the  
oxygen abundance gradients computed at 
different azimuths after 11 Gyr of disk evolution for models with spiral multiplicity
$m$ = 1 and  the same spiral pattern speeds $\Omega_{s}$ as in  Fig.
\ref{S2_grad} (see Table 1
for model details).
We notice that  around the corotation radii the azimuthal
abundance variations are generally more evident for models
with one spiral arm compared  to ones  with spiral multiplicity
$m$ = 2.

In Fig. \ref{CS1} we show
the  present day  azimuthal  residual of the
oxygen abundances after
subtracting the average radial gradient computed in annular regions
which contain the  corotation radii for 
  models with $m=1$ multiplicity: S1A, S1B, S1C, S1D, S1F (see Table 1 for other parameter
details). 
For the model S1D at the Galactic distance of  13 kpc we have
$\Delta$[O/H]$\approx$ 0.40 dex, which is about $\approx 25$\% larger than the
S2D case. As found for the model with  $m=2$, the oxygen
abundance variations become important in regions not so far from the  solar vicinity, i.e., model S1C
whose corotation resides  at $R=$ 11
kpc,  $\Delta$[O/H]  $\approx$ 0.23 dex.

\subsubsection{Results for different pitch angles}

In this Section we consider different pitch angles $\alpha$ for the
spiral arms in our Milky Way galaxy. 

Recent work by Quillen et al. (2018) and Laporte et al. (2018) suggest that tightly wound spiral structure should be considered, based on modeling of
phase-space structure found in
the second Gaia data release   (Gaia collaboration et al. 2018).

A smaller pitch angle gives rise
to more  tightly wound spiral structure.  The upper panel of  Fig. \ref{SESFSFR}  depicts the present time  SFR 
  computed with  a pitch angle
$\alpha=7^{\circ}$ (model S2G in Table 1), whereas  the lower panel shows the case  of
$\alpha=30^{\circ}$  (model S2H in Table 1). For both panels
the other model parameters as the same as model S2A. 
The spiral pattern is clearly visible in the SFR,
and for the model S2G a
tighter wound spiral structure is present. 

In Fig \ref{SE_SF} we compare the azimuthal variations for models S2G
and S2H. We see that the chemical variations are identical at the
corotation radius and simply azimuthally shifted for other Galactocentric distances.

\section{Conclusions}

In this paper we presented a new 2D chemical evolution model, able to
trace azimuthal variations in the galactic disc density. We applied
this model to (i) the density fluctuations arising in a disc formation
simulation by Martig et al. (2012), used for the MCM13 Milky Way
chemo-dynamical model, and (ii) the density perturbations originating
from an analytical spiral arm formulation.

 The main conclusions  for  density
perturbation from  Milky Way
chemo-dynamical model by MCM13 can be summarized as follows:
\begin{itemize}
\item We found that the density fluctuations produce significant oxygen azimuthal variations in
  the abundance gradients of the order of 0.1 dex.
\item The azimuthal variations are more evident in the external
  galactic regions,  in agreement with the recent observations of the
  galaxy NGC 6754, using MUSE data (S{\'a}nchez et al. 2015).
\end{itemize}

In an effort to understand the above findings, we constructed simple analytical spiral arm models, for which we varied the pattern speed, multiplicity and pitch angle with the following main findings: 
\begin{itemize}
\item 
 The larger fluctuations in the azimuthal abundance gradients are found near the corotation radius, where the relative velocity with respect to the disk is close to zero.
\item 

Larger azimuthal variations are found at corotation radii shifted to larger radii, i.e., slower pattern speeds.
\item 

The variation is more enhanced for the model with only one spiral arm, which is expected to result from the combination of an $m$=2 and $m$=3 spiral structure.

\item We found that the more significant azimuthal abundance
  variations seen at early times in  presence of a regular, periodic
  perturbation tend to quench at later times. This is expected, as galactic chemical evolution is cumulative process and phase-mixing and radial migration tends to wipe structure with time.

\end{itemize}
Combining the effect of corotaton radii by assuming the simultaneous
propagation of multiple spiral modes through galactic disks, we can
obtain a realistic picture of azimuthal variations induced at stellar
birth found in self-consistent models, such as the MCM13. Material
spiral arms propagating near the corotation at all galactic radii have
been described by a number of recent numerical work with different
interpretations (see Grand et al. 2012, Comparetta \& Quillen 2012,
Hunt et al. 2019).

In future work we will improve the new 2D chemical evolution model introduced here by taking into account stellar radial migration of long-lived stars and the pollution to the ISM abundance introduced by them at radii and azimuths different than their birth places. We will also use this model to update the Galactic habitable zone results presented by Spitoni et al. (2014, 2017) and study the effect of spiral structure and the Galactic bar.

\section*{Acknowledgement}

 We thank the anonymous referee for various suggestions
 that improved the paper.
E. Spitoni and V. Silva Aguirre acknowledge support from the Independent Research Fund Denmark (Research grant 7027-00096B). V. Silva Aguirre acknowledges support from VILLUM FONDEN (Research Grant 10118).
G. Cescutti acknowledges
financial support from the European Union
Horizon 2020 research and innovation programme
under the Marie Sklodowska-Curie grant agreement
No. 664931. This work has been partially supported
by the EU COST Action CA16117 (ChETEC).
I. Minchev acknowledges support by the Deutsche Forschungsgemeinschaft under the grant MI 2009/1-1.
F. Matteucci acknowledges research funds from the University of Trieste (FRA2016).

\end{document}